\documentclass[prd,twocolumn,pre,superscriptaddress]{revtex4-1}

\newcommand{\nue}{\ensuremath{\nu_{e}}}

\newcommand{\nuebar}{\ensuremath{\overline{\nu}_{e}}}
\newcommand{\g}{\gamma}
\newcommand{\pspt}{PROSPECT}

\newcommand{\adone}{AD-I}
\newcommand{\adtwo}{AD-II}

\newcommand{\NIST}{National Institute of Standards and Technology (NIST) \renewcommand{\NIST}{NIST}}
\newcommand{\NBSR}{National Bureau of Standards Reactor (NBSR) \renewcommand{\NBSR}{NBSR}}
\newcommand{\INL}{Idaho National Laboratory (INL) \renewcommand{\INL}{INL}}
\newcommand{\ATR}{Advanced Test Reactor (ATR) \renewcommand{\ATR}{ATR}}
\newcommand{\ORNL}{Oak Ridge National Laboratory (ORNL) \renewcommand{\ORNL}{ORNL}}
\newcommand{\HFIR}{High Flux Isotope Reactor (HFIR) \renewcommand{\HFIR}{HFIR}}
\newcommand{\LLNL}{Lawrence Livermore National Laboratory (LLNL) \renewcommand{\LLNL}{LLNL}}
\newcommand{\isot}[2]{$^{#1}$#2}
\newcommand{\ptwenty}{PROSPECT-20 (P20) \renewcommand{\ptwenty}{P20}}

\usepackage{epstopdf}
\usepackage{bm}
\usepackage[pdftex]{graphicx}
\usepackage{hyperref}
\usepackage[usenames,dvipsnames]{xcolor}
\usepackage{url}
\usepackage{tabularx}
\usepackage{textcase}
\usepackage{multirow} 
\usepackage{graphicx}
\usepackage{amssymb,amsmath}
\usepackage{bm}
\usepackage[sc]{mathpazo}
\usepackage{titlesec}
\usepackage{lineno}
\usepackage[utf8]{inputenc}

\begin{document}

\title{The PROSPECT Physics Program}

% !TEX root = main.tex

\author{J.~Ashenfelter}
\affiliation{Wright Laboratory, Department of Physics, Yale University, New Haven, CT, USA}
\author{B.~Balantekin}
\affiliation{Department of Physics, University of Wisconsin, Madison, Madison, WI, USA} 
\author{H.~R.~Band}
\affiliation{Wright Laboratory, Department of Physics, Yale University, New Haven, CT, USA}
\author{G.~Barclay}
\affiliation{High Flux Isotope Reactor, Oak Ridge National Laboratory, Oak Ridge, TN, USA}
\author{C.~D.~Bass}
\affiliation{Department of Chemistry and Physics, Le Moyne College, Syracuse, NY, USA }
\author{D.~Berish}
\affiliation{Department of Physics, Temple University, Philadelphia, PA, USA}
\author{N.~S.~Bowden}
\affiliation{Nuclear and Chemical Sciences Division, Lawrence Livermore National Laboratory, Livermore, CA, USA}
\author{A.~Bowes}
\affiliation{Department of Physics, Illinois Institute of Technology, Chicago, IL, USA}
\author{C.~D.~Bryan}
\affiliation{High Flux Isotope Reactor, Oak Ridge National Laboratory, Oak Ridge, TN, USA}
\author{J.~P.~Brodsky}
\affiliation{Nuclear and Chemical Sciences Division, Lawrence Livermore National Laboratory, Livermore, CA, USA}
\author{J.~J.~Cherwinka}
\affiliation{Physical Sciences Laboratory, University of Wisconsin, Madison, Madison, WI, USA}
\author{R.~Chu}
\affiliation{Physics Division, Oak Ridge National Laboratory, Oak Ridge, TN, USA}
\affiliation{Department of Physics and Astronomy, University of Tennessee, Knoxville, TN, USA}
\author{T.~Classen}
\affiliation{Nuclear and Chemical Sciences Division, Lawrence Livermore National Laboratory, Livermore, CA, USA}
\author{K.~Commeford}
\affiliation{Department of Physics, Drexel University, Philadelphia, PA, USA}
\author{D.~Davee}
\affiliation{Department of Physics, College of William and Mary, Williamsburg, VA, USA}
\author{D.~Dean}
\affiliation{Physics Division, Oak Ridge National Laboratory, Oak Ridge, TN, USA}
\author{G.~Deichert}
\affiliation{High Flux Isotope Reactor, Oak Ridge National Laboratory, Oak Ridge, TN, USA}
\author{M.~V.~Diwan}
\affiliation{Physics Department, Brookhaven National Laboratory, Upton, NY, USA}
\author{M.~J.~Dolinski}
\affiliation{Department of Physics, Drexel University, Philadelphia, PA, USA}
\author{J.~Dolph}
\affiliation{Physics Department, Brookhaven National Laboratory, Upton, NY, USA}
\author{J.~K.~Gaison}
\affiliation{Wright Laboratory, Department of Physics, Yale University, New Haven, CT, USA}
\author{A.~Galindo-Uribarri}
\affiliation{Physics Division, Oak Ridge National Laboratory, Oak Ridge, TN, USA}
\affiliation{Department of Physics and Astronomy, University of Tennessee, Knoxville, TN, USA}
\author{K.~Gilje}
\affiliation{Department of Physics, Illinois Institute of Technology, Chicago, IL, USA}
\author{A.~Glenn}
\affiliation{Nuclear and Chemical Sciences Division, Lawrence Livermore National Laboratory, Livermore, CA, USA}
\author{B.~W.~Goddard}
\affiliation{Department of Physics, Drexel University, Philadelphia, PA, USA}
\author{M.~Green}
\affiliation{Physics Division, Oak Ridge National Laboratory, Oak Ridge, TN, USA}
\author{K.~Han}
\affiliation{Wright Laboratory, Department of Physics, Yale University, New Haven, CT, USA}
\author{S.~Hans}
\affiliation{Chemistry Department, Brookhaven National Laboratory, Upton, NY, USA}
\author{K.~M.~Heeger}
\affiliation{Wright Laboratory, Department of Physics, Yale University, New Haven, CT, USA}
\author{B.~Heffron}
\affiliation{Physics Division, Oak Ridge National Laboratory, Oak Ridge, TN, USA}
\affiliation{Department of Physics and Astronomy, University of Tennessee, Knoxville, TN, USA}
\author{D.~E.~Jaffe}
\affiliation{Physics Department, Brookhaven National Laboratory, Upton, NY, USA}
\author{D.~Jones}
\affiliation{Department of Physics, Temple University, Philadelphia, PA, USA}
\author{T.~J.~Langford}
\affiliation{Wright Laboratory, Department of Physics, Yale University, New Haven, CT, USA}
\author{B.~R.~Littlejohn}
\affiliation{Department of Physics, Illinois Institute of Technology, Chicago, IL, USA}
\author{D.~A.~Martinez~Caicedo}
\affiliation{Department of Physics, Illinois Institute of Technology, Chicago, IL, USA}
\author{R.~D.~McKeown}
\affiliation{Department of Physics, College of William and Mary, Williamsburg, VA, USA}
\author{M.~P.~Mendenhall}
\affiliation{National Institute of Standards and Technology, Gaithersburg, MD, USA}
\author{P.~Mueller}
\affiliation{Physics Division, Oak Ridge National Laboratory, Oak Ridge, TN, USA}
\author{H.~P.~Mumm}
\affiliation{National Institute of Standards and Technology, Gaithersburg, MD, USA}
\author{J.~Napolitano}
\affiliation{Department of Physics, Temple University, Philadelphia, PA, USA}
\author{R.~Neilson}
\affiliation{Department of Physics, Drexel University, Philadelphia, PA, USA}
\author{D.~Norcini}
\affiliation{Wright Laboratory, Department of Physics, Yale University, New Haven, CT, USA}
\author{D.~Pushin}
\affiliation{Institute for Quantum Computing and Department of Physics, University of Waterloo, Waterloo, ON, Canada}
\author{X.~Qian}
\affiliation{Physics Department, Brookhaven National Laboratory, Upton, NY, USA}
\author{E.~Romero}
\affiliation{Physics Division, Oak Ridge National Laboratory, Oak Ridge, TN, USA}
\affiliation{Department of Physics and Astronomy, University of Tennessee, Knoxville, TN, USA}
\author{R.~Rosero}
\affiliation{Chemistry Department, Brookhaven National Laboratory, Upton, NY, USA}
\author{B.~S.~Seilhan}
\affiliation{Nuclear and Chemical Sciences Division, Lawrence Livermore National Laboratory, Livermore, CA, USA}
\author{R.~Sharma}
\affiliation{Physics Department, Brookhaven National Laboratory, Upton, NY, USA}
\author{S.~Sheets}
\affiliation{Nuclear and Chemical Sciences Division, Lawrence Livermore National Laboratory, Livermore, CA, USA}
\author{P.~T.~Surukuchi}
\affiliation{Department of Physics, Illinois Institute of Technology, Chicago, IL, USA}
\author{R.~L.~Varner}
\affiliation{Physics Division, Oak Ridge National Laboratory, Oak Ridge, TN, USA}
\author{B.~Viren}
\affiliation{Physics Department, Brookhaven National Laboratory, Upton, NY, USA}
\author{W.~Wang}
\affiliation{School of Physics and Engineering, Sun Yat-Sen University, Guangzhou, Guangdong Province, China}
\affiliation{Department of Physics, College of William and Mary, Williamsburg, VA, USA}
\author{B.~White}
\affiliation{Physics Division, Oak Ridge National Laboratory, Oak Ridge, TN, USA}
\author{C.~White}
\affiliation{Department of Physics, Illinois Institute of Technology, Chicago, IL, USA}
\author{J.~Wilhelmi}
\affiliation{Department of Physics, Temple University, Philadelphia, PA, USA}
\author{C.~Williams}
\affiliation{Physics Division, Oak Ridge National Laboratory, Oak Ridge, TN, USA}
\author{T.~Wise}
\affiliation{Wright Laboratory, Department of Physics, Yale University, New Haven, CT, USA}
\author{H.~Yao}
\affiliation{Department of Physics, College of William and Mary, Williamsburg, VA, USA}
\author{M.~Yeh}
\affiliation{Chemistry Department, Brookhaven National Laboratory, Upton, NY, USA}
\author{Y.-R.~Yen}
\affiliation{Department of Physics, Drexel University, Philadelphia, PA, USA}
\author{G.~Zangakis}
\affiliation{Department of Physics, Temple University, Philadelphia, PA, USA}
\author{C.~Zhang}
\affiliation{Physics Department, Brookhaven National Laboratory, Upton, NY, USA}
\author{X.~Zhang}
\affiliation{Department of Physics, Illinois Institute of Technology, Chicago, IL, USA}

\collaboration{The PROSPECT Collaboration}

\begin{abstract}
The Precision Reactor Oscillation and Spectrum Experiment, PROSPECT, is designed to make a precise measurement of the antineutrino spectrum from a highly-enriched uranium reactor and probe eV-scale sterile neutrinos by searching for neutrino oscillations over meter-long distances. PROSPECT is conceived as a 2-phase experiment utilizing segmented $^6$Li-doped liquid scintillator detectors for both efficient detection of reactor antineutrinos through the inverse beta decay reaction and excellent background discrimination. PROSPECT Phase~I consists of a movable 3-ton antineutrino detector at distances of 7--12 m from the reactor core. It will probe the best-fit point of the \nue{} disappearance experiments at 4$\sigma$ in 1 year and the favored region of the sterile neutrino parameter space at $>$3$\sigma$ in 3 years. With a second antineutrino detector at 15--19~m from the reactor, Phase~II of PROSPECT can probe the entire allowed parameter space below 10~eV$^{2}$ at 5$\sigma$ in 3 additional years. The measurement of the reactor antineutrino spectrum and the search for short-baseline oscillations with PROSPECT will test the origin of the spectral deviations observed in recent $\theta_{13}$ experiments, search for sterile neutrinos, and conclusively address the hypothesis of sterile neutrinos as an explanation of the reactor anomaly. 
\end{abstract}

\maketitle

\tableofcontents

% !TEX root = main.tex

\section{Executive Summary and Background}
\label{sec:intro}

\begin{figure}[b]
   \centering
    \includegraphics[width=0.48\textwidth]{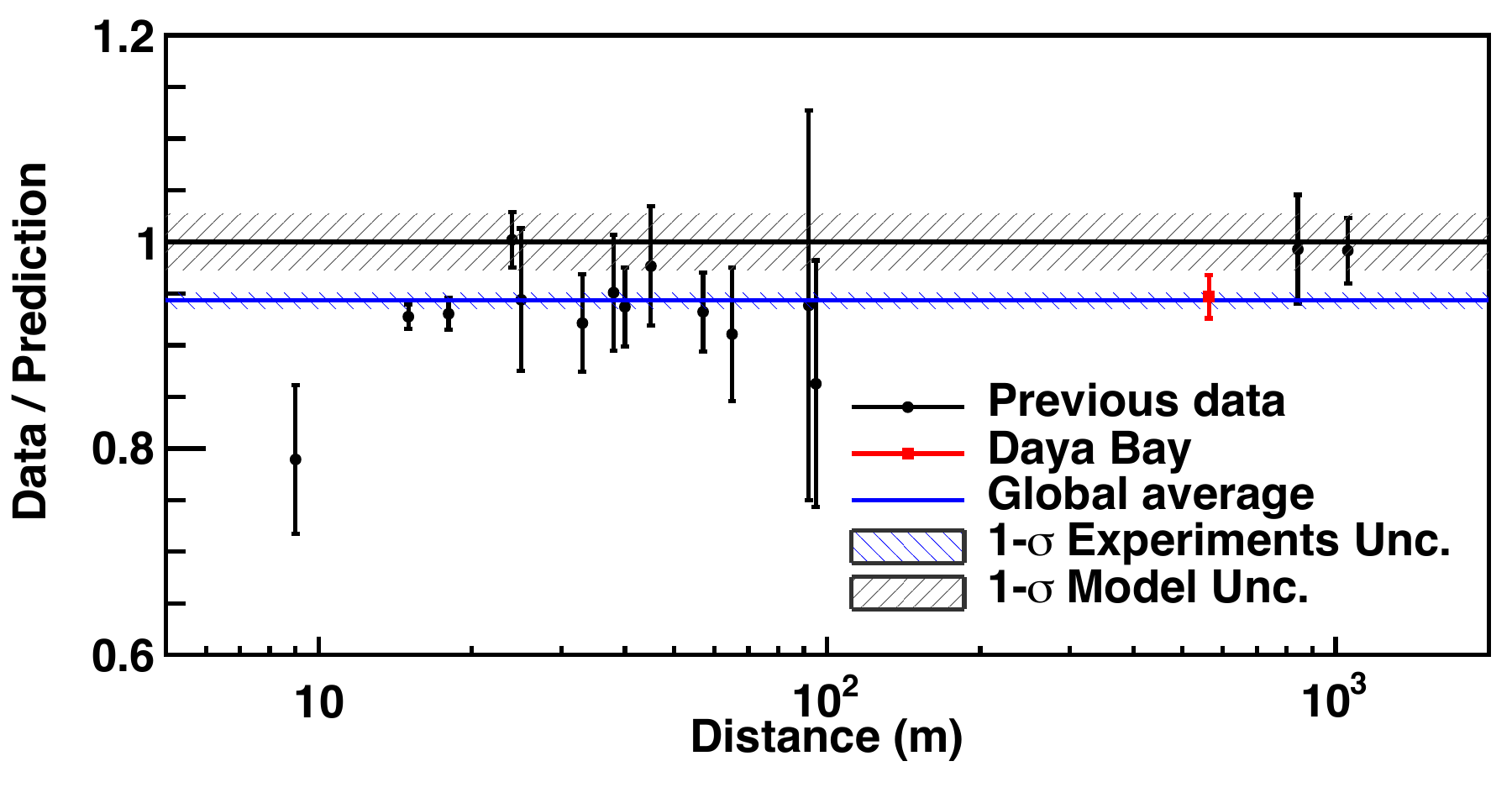}
   \caption{Comparison of measured reactor antineutrino fluxes with predictions based on models for the emission of reactor antineutrinos. The measured deficit relative to prediction is known as the ``reactor antineutrino anomaly''~\cite{Mention:2011rk}. Fig.~1 from~\cite{An:2015nua}.}
   \label{fig:reactorAnomaly}
\end{figure}

\begin{figure}[b]
   \centering
    \includegraphics[clip=true, trim=0mm 0mm 0mm 0mm,width=0.43\textwidth]{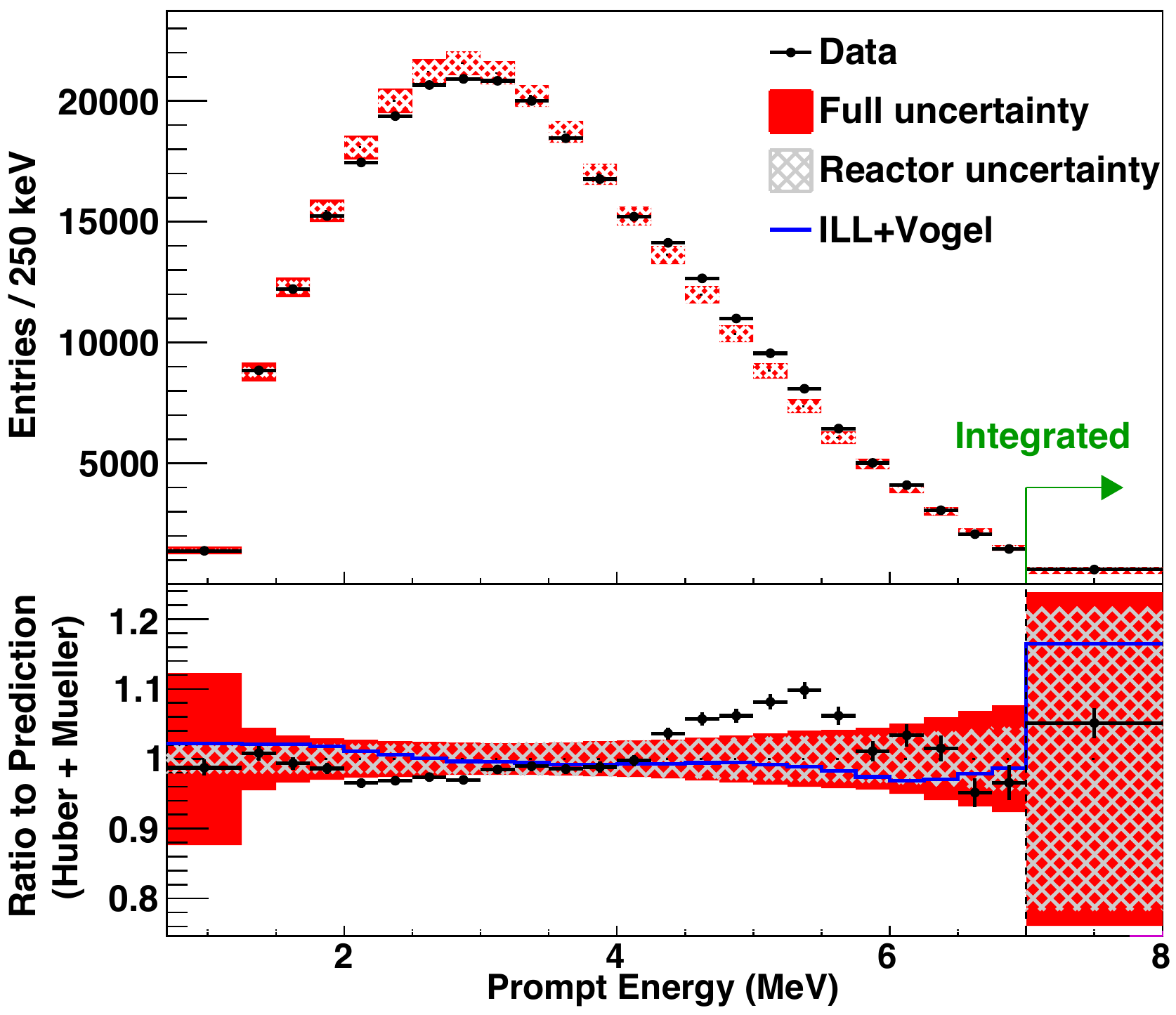}
  \caption{Comparison of the measured antineutrino energy spectra from pressure water reactors (PWR) to modern models for reactor neutrino emission observed in the Daya Bay experiment (Fig.~2 from~\cite{An:2015nua}). The deviation from prediction between 4--6~MeV, which is also observed in two other similar experiments, is unexplained and may indicate deficiencies in the models and/or the nuclear data underlying them.}
   \label{fig:spectrumAnomaly}
\end{figure}

Recent neutrino experiments have provided a coherent picture of neutrino flavor change and mixing and allowed the precise determination of oscillation parameters in the 3-neutrino model. However, anomalous results in the measurement of the reactor \nuebar{} spectrum and flux have suggested this picture is incomplete and may be interpreted as indicators of new physics.  
Reactor \nuebar{} experiments observe a $\sim$6\% deficit in the absolute flux when compared to predictions~\cite{Mueller:2011nm,Huber:2011wv}. 
The observed flux deficit, the ``reactor antineutrino anomaly'', has led to the hypothesis of oscillations involving a sterile neutrino state with $\sim$1~eV$^2$ mass splitting~\cite{Mention:2011rk,Abazajian:2012ys,Kopp:2013vaa} (Fig.~\ref{fig:reactorAnomaly}). 
Moreover, measurements of the reactor \nuebar{} spectrum by recent $\theta_{13}$ experiments indicate spectral discrepancies compared to predictions, particularly at \nuebar{} energies of 4--7~MeV~\cite{An:2015nua,Abe:2014bwa,Kim:2014rfa} (Fig.~\ref{fig:spectrumAnomaly}). 
The reactor anomaly and the measured spectral deviations are open issues in a suite of anomalous results~\cite{Abazajian:2012ys} that may hint at revolutionary new physics in the neutrino sector. 

The Precision Reactor Oscillation and Spectrum Experiment, PROSPECT \cite{prospectweb}, is designed to comprehensively address this situation by making a definitive search for \nuebar{} oscillations at short baselines from a compact reactor core while concurrently making the world's most precise \nuebar{} energy spectrum measurement from a highly-enriched uranium (HEU) research reactor.   
By simultaneously measuring the relative \nuebar{} flux and spectrum at multiple distances from the core within the same detector, PROSPECT will probe for oscillations into additional neutrino states in the parameter space favored by reactor and radioactive source experiments independent of any reactor model predictions. 
PROSPECT covers a unique region of parameter space at the eV$^2$-scale that is complementary to current \nuebar{} disappearance limits from Daya Bay and to $\nu_{\mu}$ and \nue{} oscillation searches in accelerator-based neutrino oscillation experiments.  
Together, reactor and accelerator experiments define a comprehensive approach to resolving the current anomalous results in neutrino physics  and have the potential to make a paradigm-changing discovery.  
Observation of an eV-scale sterile neutrino would have a profound impact on our understanding of neutrino physics and the Standard Model of particle physics with wide-ranging implications for the physics reach of the planned US long-baseline experiment DUNE~\cite{Gandhi:2015xza}, searches for neutrinoless double beta decay, neutrino mass constraints from cosmology and beyond.

\begin{figure*}[t]
  \begin{center}
    \includegraphics[ trim=0.0cm 0.0cm 0.0cm 0.0cm, clip=true,width=0.6\textwidth]{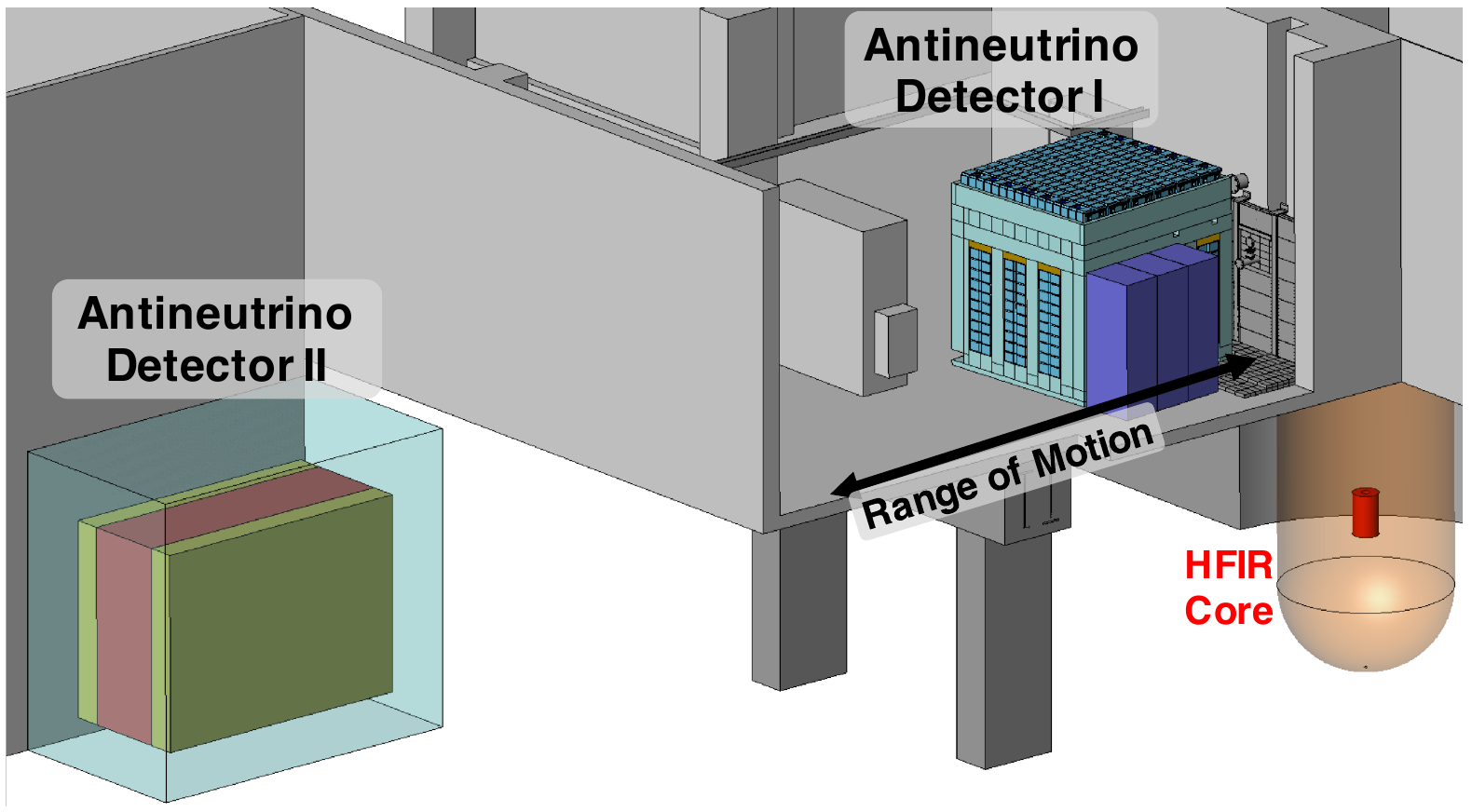}
  \end{center}
\caption{
Layout of the PROSPECT experiment. Shown are the HFIR reactor core and the two antineutrino detectors, \adone{} and \adtwo{}. Phase~I consists of a movable antineutrino detector, \adone{}, operated for three years at a baseline range of 7--12~m. Phase~II adds a $\sim$10-ton detector, \adtwo{}, at 15--19~m for an extra three years of data taking.}  
\label{fig:experimentalLayout}
\end{figure*}

By making a high-resolution energy spectrum measurement, PROSPECT will also determine if the observed spectral deviations in Daya Bay and other $\theta_{13}$ experiments at commercial nuclear power plants persist in a HEU research reactor and provide a precision benchmark spectrum to test the modeling of reactor \nuebar{} production. 
A better understanding of the reactor \nuebar{} spectrum will aid precision medium-baseline reactor experiments such as JUNO and RENO-50 \cite{Capozzi:2015bpa}, and improve reactor monitoring capabilities for nonproliferation and safeguards. 

The goals of the \pspt{} experiment are to:
\begin{itemize}
\item Make an unambiguous discovery of eV-scale sterile neutrinos through the observations of energy and baseline dependent oscillation effects, or exclude the existence of this particle in the allowed parameter region with high significance. Accomplishing this addresses the proposed sterile neutrino explanation of the reactor anomaly using a method that is independent of reactor flux predictions;
\item Directly test reactor antineutrino spectrum predictions using a well-understood and modeled reactor system, while also providing information that is complementary to nuclear data measurement efforts;
\item Demonstrate techniques for antineutrino detection on the surface with little overburden;
\item Develop technology for use in nonproliferation applications.
\end{itemize}

PROSPECT will employ a phased approach for the timely exploration of the favored parameter space with the potential of a high-impact discovery while preparing for a definitive measurement across the entire allowed parameter region. PROSPECT will be located at the \HFIR{}~\cite{HFIR} at \ORNL{}~\cite{ORNL}. The proposed layout of the experiment is shown in Fig.~\ref{fig:experimentalLayout}. Phase~I of PROSPECT consists of a $\sim$3-ton, segmented $^6$Li-doped liquid scintillator antineutrino detector (\adone{}) accessing baselines in the range 7--12~m from the reactor core.  
Phase~II involves the addition of a second $\sim$10-ton antineutrino detector (\adtwo{}) with identical segmentation spanning baselines between 15--19~m.  

PROSPECT combines competitive exposure, baseline mobility for increased physics reach and systematic checks, good energy and position resolution, and efficient background discrimination that yields a signal over background sufficient to achieve stated goals. 
The additional second detector in Phase~II will allow PROSPECT to become the most sensitive experiment of all proposed reactor-based short-baseline searches.
Within a single calendar year, \pspt{} Phase I can probe the best-fit region for all current global analyses of \nue{} and \nuebar{} disappearance~\cite{Abazajian:2012ys,Kopp:2013vaa} at 4$\sigma$ confidence level.
Over 3 years of operation, \pspt{} Phase~I can discover oscillations as a sign of sterile neutrinos at 5$\sigma$ for the best-fit point and 3$\sigma$ over the majority of the suggested parameter space. 
After 3 additional years of operation with a second antineutrino detector in Phase~II, essentially all parameter space suggested by $\nu_e$ disappearance data below 10~eV$^2$ can be excluded at 5$\sigma$.

% !TEX root = main.tex

\subsection{The Reactor Antineutrino Anomaly}
\label{sec:reactorAnomaly}

Recent atmospheric, solar, reactor, and accelerator experiments have established the framework of neutrino mixing and flavor change and determined related neutrino oscillation parameters.  
Reactor experiments have utilized the inverse beta decay interaction to detect \nuebar{} emitted by beta decays of   fission daughter products and measure the flux and spectrum of reactor  \nuebar{} over a range of distances from reactors. 
Prior to the discovery of neutrino oscillations, experiments positioned $<$100~m from a variety of reactor cores, including those at ILL-Grenoble, Bugey, and Savannah River,  measured the flux of \nuebar{} with ton-scale detectors based on liquid scintillators and/or $^3$He proportional counters~\cite{ILL:1981ua,Goesgen:1986cu,Krasnoyarsk1:1987ue, Rovno88:1988gx,Rovno91:1990ry,Declais:1994su,Bugey4:1994ma,Krasnoyarsk2:1994ut, SRP:1996pb}.
These results were in good agreement with contemporary predictions based on conversion of  $\beta^-$ spectra of fissioning isotopes measured at the ILL-Grenoble research reactor~\cite{Schreckenbach:1985ep, Hahn:1989zr}. 

Motivated by experiments seeking to measure $\theta_{13}$, an improved prediction of the reactor \nuebar{} flux was performed~\cite{Mueller:2011nm} using a novel approach combining \textit{ab-initio} and conversion methods, incorporating updated nuclear data, and more accurate nuclear corrections. 
The summation, or \textit{ab-initio}, portion of the prediction built the \nuebar{} spectrum from the sum of daughter products contributions for which fission yield, branching ratio, and decay information were available from nuclear databases, allowing nuclear corrections to be applied at the branch level. 
The residual $\sim$10\% of the spectrum was derived via a conversion procedure using the reference ILL $\beta^-$ spectrum~\cite{Schreckenbach:1981,VonFeilitzsch:1982jw,Schreckenbach:1985ep,Hahn:1989zr}. The residual contribution to the total $\beta^-$ spectra from fission daughter products isotopes without nuclear data was fit using five virtual $\beta$-branches, where importantly and in contrast to past prediction methods, corrections were applied at the branch level. The \nuebar{} spectrum obtained following this approach, when combined with the inverse beta decay cross section,  resulted in a systematic increase in the detectable reactor \nuebar{} flux. % of 3.5\%. 
Note that this increase %in detected flux 
is due to the improved evaluation %of the spectrum 
resulting in a greater proportion of the emitted \nuebar{} being above the threshold for inverse beta decay -- the total  \nuebar{} flux is still anchored to the  normalization of the ILL $\beta^-$ measurement.
In conjunction with the revision of the neutron mean lifetime~\cite{PDG2014}, this effect results in an average deficit of 5.7\% in all the short-baseline reactor \nuebar{} measurements. 
This discrepancy between data and prediction, referred to as the ``reactor antineutrino anomaly''~\cite{Mention:2011rk}, represents a deficit in the ratio of observed to expected \nuebar{} from unity significant at 98.6\% confidence level.

An independent cross-check was performed using an approach based only on an improved conversion of the ILL reference $\beta^-$ spectrum, which minimized the use of nuclear databases~\cite{Huber:2011wv}. 
Virtual $\beta$-branches were used to convert the ILL reference to an \nuebar{} spectrum, yielding a net increase of $\sim$6$\%$ in antineutrino predictions, consistent with the flux predicted in~\cite{Mueller:2011nm}. 
The cause of the increase relative to past predictions was also understood to be due to the use of improved nuclear corrections, the updated neutron lifetime, and the application of corrections to the beta decay spectrum at the branch level, in contrast to the ``effective'' correction used in past predictions. Additionally, blind analyses from recent kilometer baseline precision rate measurements are consistent with the previous reactor experiments ~\cite{An:2015nua,Abe:2014bwa,Kim:2014rfa}. The disagreement between modern reactor \nuebar{} flux predictions and measurement is therefore well-established. 

Oscillations at short baselines due to a new type of neutrino with a mass splitting of $\Delta m^{2}$$\sim$1~eV$^{2}$ have been proposed as one explanation for these observations~\cite{Mention:2011rk}.
With invisible decay width results from Z boson measurements stringently limiting the number of active neutrino flavors to three~\cite{PDG2014}, any additional existing neutrino should be 'sterile' and not participate in weak interactions.
The oscillation arising from such a neutrino with eV-scale mass splitting can be observed at baselines around 10~m from a compact reactor core.

Deficiencies in the flux prediction methods and/or imperfections in the measured data underlying them should also be considered as an explanation for the ``reactor anomaly.'' 
For example, incomplete nuclear data for the beta decays contributing to the reactor spectrum as well as uncertainties in the corrections applied to individual beta spectra may lead to significant uncertainties in the conversion procedure between the reference beta electron  and the emitted \nuebar{} spectra~\cite{Hayes:2013wra}. 
Observed spectral discrepancies in addition to the flux deficit, as described in the next section, highlight this concern. 
More data is needed to understand and explain these observations.  
\pspt{} can address both of these possibilities through a high precision spectral measurement in addition to an oscillation search for sterile neutrinos, and therefore provide a comprehensive solution to the present ``reactor anomalies.''

% !TEX root = main.tex

\subsection{Reactor Spectrum Anomaly}

Neutron-rich fission fragments within a reactor emit \nuebar{} via beta decay with an energy spectrum dependent on the transition between initial and final nuclear states of the particular isotope.  
The total energy spectrum $S(E_{\overline{\nu}})$ can be expressed as a sum of the decay rate of each unstable isotope $i$ in the reactor, $R_i$, times the branching fraction for beta decay $f_{ij}$ to the nuclear state $j$ with an energy spectrum $S_{ij}(E_{\overline{\nu}})$,
\begin{equation}
  S(E_{\overline{\nu}}) = \sum_i R_i\sum_j f_{ij}S_{ij}(E_{\overline{\nu}}).
\end{equation}
While this calculation is straightforward in principle, it is complex in practice.  
More than 1000 unstable isotopes contribute, and many fission yields and individual beta decay spectra are poorly known.  
For those measured, there can still be significant uncertainty in the decay levels, branching fractions, and $\overline{\nu}_e$ energy spectra.  
It has recently been demonstrated that the two major nuclear databases, ENDF and JEFF, contain differences in branching fractions~\cite{Hayes:2015yka}, complicating the interpretation of these calculations. 
Separately, total-absorption gamma spectroscopy measurements of key isotopes have shown that quoted uncertainties are frequently underestimated~\cite{Zakari-Issoufou:2015vvp}.
Consequently, \textit{ab-initio} calculations of $S(E_{\overline{\nu}})$ are thought accurate to only $\sim$10\%~\cite{Hayes:2013wra}.

Given the uncertainties in this approach, the conversion method has become the de-facto standard for modeling reactor \nuebar{} energy spectra. 
The cumulative $\beta^-$ energy spectra emitted by foils of fissioning material were measured~\cite{Schreckenbach:1981,VonFeilitzsch:1982jw, Schreckenbach:1985ep, Hahn:1989zr,Haag:2013raa} and used to estimate the corresponding cumulative \nuebar{} spectra with an  estimated uncertainty at the few-percent level.
As described in Sec.~\ref{sec:reactorAnomaly} however, modern predictions of this type disagree with measurements of the flux, leading to the reactor antineutrino anomaly.   In addition, recent, high-precision measurements of the antineutrino energy spectrum from $\theta_{13}$ experiments have shown deviations from the theoretically predicted spectral shapes.
The measured spectra from Daya Bay, Double Chooz, and RENO each show an excess of antineutrinos of approximately 10\% with energies between 5 and 7~MeV~\cite{An:2015nua,Abe:2014bwa,Kim:2014rfa}.   

Initial studies indicated that the \textit{ab-initio} method reproduced the shape of the spectrum better than the beta-conversion predictions~\cite{Dwyer:2014eka}.
However, re-analyses with updated fission and beta-branch information call this result into question and instead point to antineutrinos produced by the $^{238}$U fission chain as a possible source of the spectral anomaly~\cite{Hayes:2015yka}.  
New measurements with total-absorption gamma spectrometers at ORNL~\cite{Rasco:2015ufa} and University of Jyv\"{a}skyl\"{a}~\cite{Zakari-Issoufou:2015vvp} will reduce uncertainties in individual beta-decay levels and branching ratios.
However, predicting antineutrino spectra resulting from these decays remains challenging due to unknown shape corrections.
Similarly, uncertainties in the cumulative fission yields are not addressed by these measurements.
Precision measurements of reactor antineutrino spectra provide a unique experimental probe that can address many of these questions~\cite{Hayes:2015yka}.
In particular, a first-ever precision measurement of the $^{235}$U spectrum would highly constrain predictions for a  static single fissile isotope system, as compared to commercial power reactors that have evolving fuel mixtures of multiple fissile isotopes.

%!TEX root = main.tex
\subsection{Anomalies in Source and Accelerator Experiments}
\label{sec:acceleratorAnomaly}

Anomalous results have also been obtained in other neutrino experiments. Both the SAGE and GALLEX radiochemical experiments have observed neutrino flux deficits with high-activity \nue{} calibration sources~\cite{Anselmann:1994ar, Hampel:1997fc, Abdurashitov:1998ne, Abdurashitov:2005tb}.  

Additional anomalies have become apparent in accelerator-based neutrino experiments.  The Liquid Scintillator Neutrino Detector (LSND) Experiment at Los Alamos National Laboratory was designed to search for neutrino oscillations in the $\overline{\nu}_{\mu} \rightarrow \nuebar{}$ channel.  It measured an excess of events at low energy consistent with an oscillation mass splitting of $|\Delta m^{2}|$$\sim$1~eV$^{2}$~\cite{Aguilar:2001ty}.  The Mini Booster Neutrino Experiment (MiniBooNE) at Fermilab was conceived to test this so-called ``LSND anomaly'' in the same L/E region~\cite{Aguilar-Arevalo:2013pmq}.  In both the $\overline{\nu}_{\mu} \rightarrow \nuebar{}$ and $\nu_{\mu} \rightarrow \nu_e$ appearance channels, it observed an excess of events. There is some disagreement regarding the compatibility of MiniBooNE \nuebar{} appearance data in models involving 3 active neutrinos and 1 sterile state (3+1 model)~\cite{Giunti:2013aea} but the allowed regions for neutrino oscillations partially overlap with the allowed regions from LSND. 

\begin{figure}[tbp]
   \centering
    \includegraphics[trim={0cm 0cm 2cm 0cm},clip,width=0.45\textwidth]{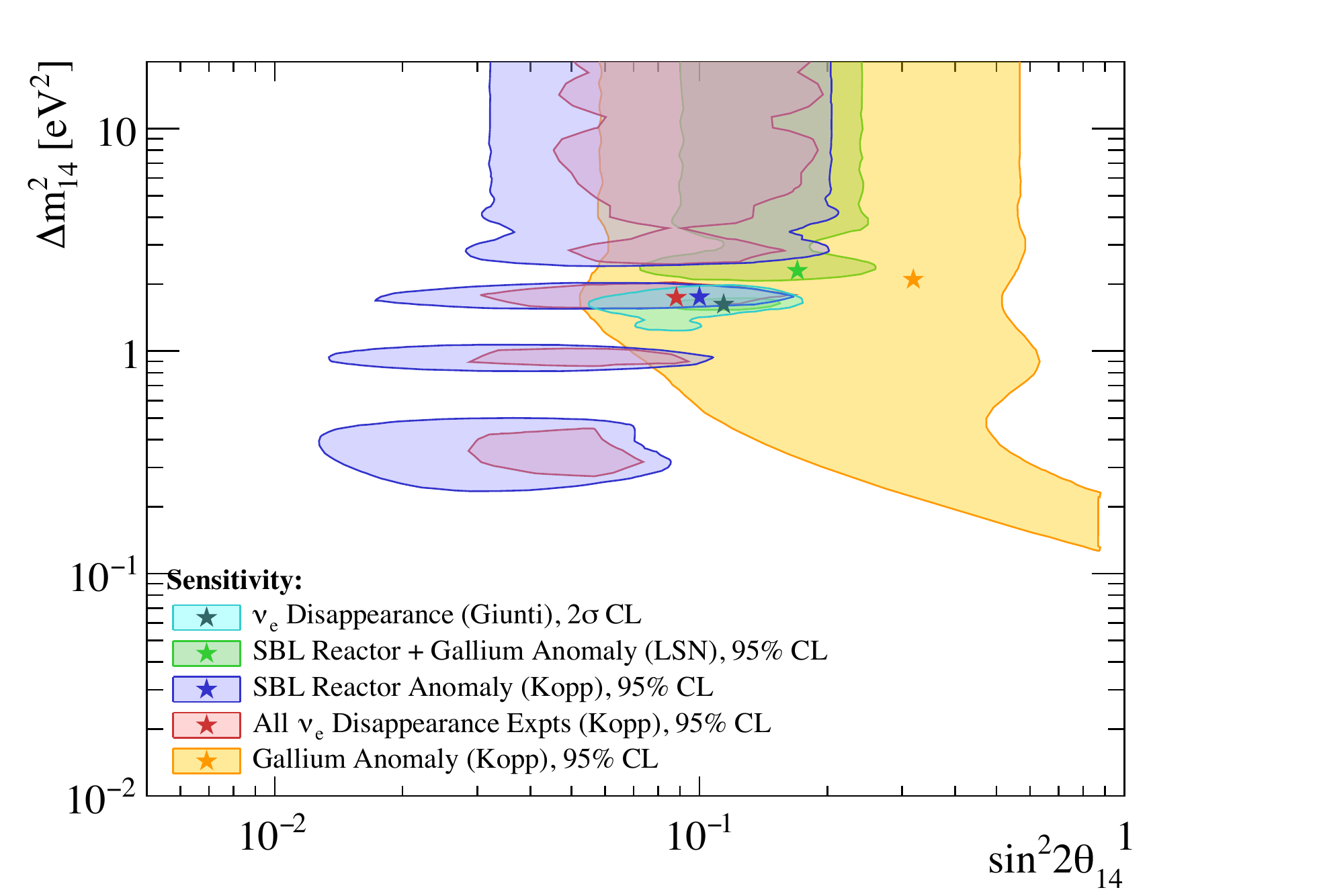}
   \caption{Allowed regions in 3+1 framework  for several combinations of \nue{} and \nuebar{} disappearance experiments. Contours obtained from ~\cite{Kopp:2013vaa,Mention:2011rk,Giunti:2013aea}.} 
   \label{fig:SBN_dis}
\end{figure}

\subsection{Global Fits}

Attempts have been made to fully incorporate the observed anomalies into a 3+1 framework of neutrino oscillations.
Combining the short-baseline reactor anomaly data with the gallium measurements under the assumption of one additional sterile neutrino state allows one to determine the allowed regions ($\Delta$m$_{14}^2$, $\sin^{2}2\theta_{14}$) in the global parameter space. 
Two recent efforts obtain slightly different allowed regions and global best-fit points~\cite{Kopp:2013vaa,Mention:2011rk}. 
The disagreement can be attributed to the differences in handling uncertainties and the choice of spectral information included in the analyses. 
Inclusion of all \nue{} and \nuebar{} disappearance measurements further constrains the parameter space~\cite{Kopp:2013vaa}. 
Fig.~\ref{fig:SBN_dis} illustrates the allowed regions obtained from different combinations of anomalous experimental results. 

Because of the tensions between some appearance and disappearance results, difficulties arise in developing a consistent picture of oscillations in the 3+1 framework~\cite{Giunti:2013aea} involving both appearance and disappearance data.
Efforts at performing a global fit in frameworks containing two additional sterile neutrinos have  produced results that have only slightly better compatibility~\cite{Kopp:2013vaa}. 
Excluding the MiniBooNE low-energy excesses yields allowed regions in a 3+1 framework and has been suggested~\cite{Giunti:2013aea} as a ``pragmatic approach,'' based on the observation that the MiniBooNE anomalies cannot be explained in any of the frameworks. 

A short-baseline oscillation experiment sensitive to the phase space in $\Delta$m$_{14}^2$ and $\sin^{2}2\theta_{14}$ 
suggested by \nue{} and \nuebar{} global fits (see Fig.~\ref{fig:SBN_dis}) will be to able to conclusively address the $\sim$1~eV$^{2}$ sterile neutrino interpretation of these anomalous results.

\subsection{Implications for the Future Neutrino Program}
The discovery of sterile neutrinos would be a profound result with far-reaching implications for nuclear and particle physics as well as cosmology. Sterile neutrinos would represent a new class of particles and possibly indicate non-standard interactions. An extension of the Standard Model would be required and our fundamental understanding of particles and interactions would change. The addition of sterile states would require an extension of the PMNS matrix describing neutrino mixing and have a dramatic impact on our interpretation of future neutrino experiments. 

The Fermilab Long-Baseline Neutrino Facility (LBNF) and the Deep Underground Neutrino Experiment (DUNE) aim to make a precision measurement of neutrino oscillation with the goal of determining the mass hierarchy and the CP-violating phase. The existence of a single eV-scale sterile neutrino implies an additional mass eigenstate, which leads to 3 new mass splittings, mixing angles, and CP-violating phases that can alter the oscillation behavior of neutrinos over the 1300~km DUNE baseline. One or more sterile neutrino states would significantly impact the interpretation of the neutrino event rate and spectrum measured in DUNE. 

\begin{figure}[tbp]
   \centering
    \includegraphics[clip=true, width=0.49\textwidth]{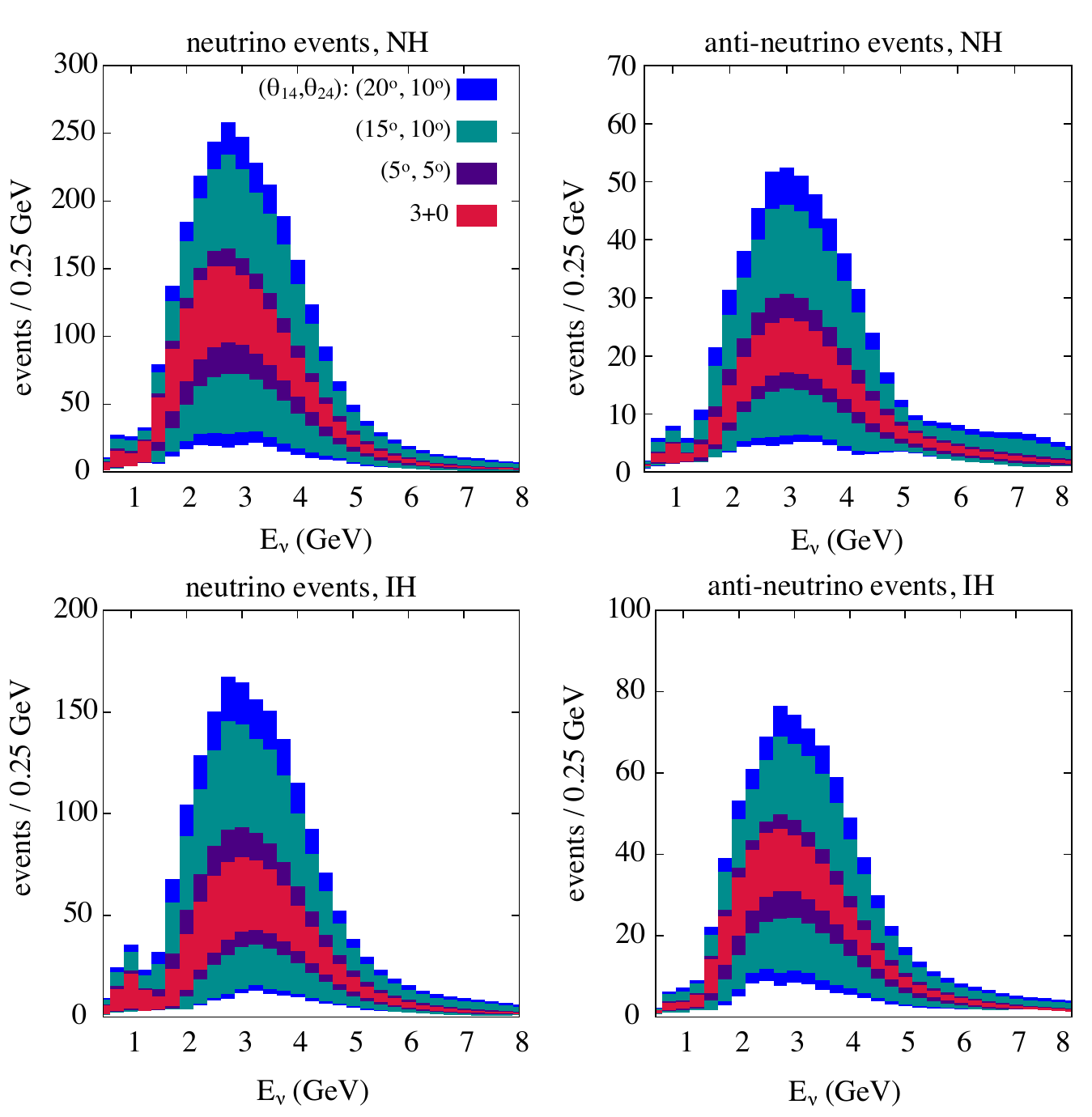}
   \caption{Expected neutrino and antineutrino rates for DUNE as a function of reconstructed neutrino energy with and without sterile neutrinos, from \cite{Gandhi:2015xza}. Each band corresponds to the range of event rates possible for a particular case. In the 3+0 scenario labelled by the red band, $\delta_{CP}$ is varied from [-180$^{\circ}$, 180$^{\circ}$]. For the 3+1 case, three sets of ($\theta_{14}$, $\theta_{24}$) are represented: blue band (20$^{\circ}$, 10$^{\circ}$), green band (15$^{\circ}$, 10$^{\circ}$), purple band (5$^{\circ}$, 5$^{\circ}$). Each set varies $\theta_{34}$ from [0$^{\circ}$, 30$^{\circ}$] and CP-violating phases $\delta_{13}$, $\delta_{24}$, and $\delta_{34}$ from [-180$^{\circ}$, 180$^{\circ}$]. }
   \label{fig:sterile_dune_LBL}
\end{figure}

Fig.~\ref{fig:sterile_dune_LBL} shows the expected rates of neutrinos and antineutrinos for DUNE in the standard three neutrino framework (3+0) and a 3+1 framework for three different combinations of mixing angles $\theta_{14}$ and $\theta_{24}$ and allowing the 3 CP-violating phases to vary from [-180$^{\circ}$, 180$^{\circ}$]. 
The difference in event rates between the 3+0 and 3+1 cases can lead to ambiguities in the interpretation of CP-violation searches and complicate the precision measurement of neutrino parameters in the DUNE experiment.

\begin{figure}[tbp]
   \centering
      \includegraphics[width=0.49\textwidth]{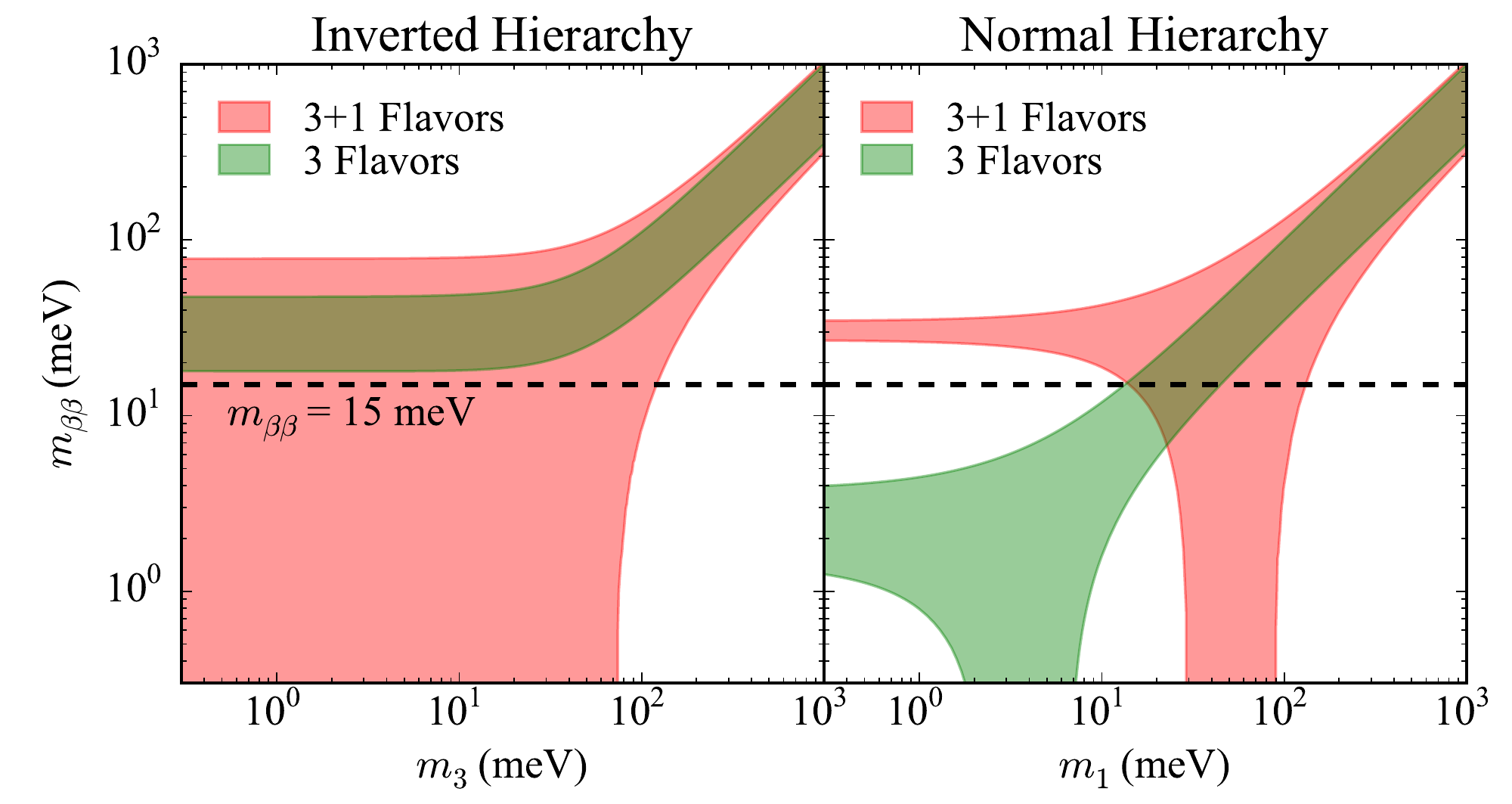}
\caption{The influence of sterile neutrinos on  the effective neutrino mass $m_{\beta\beta}$ accessible in neutrinoless double beta decay experiments for inverted hierarchy (left) and normal hierarchy (right) scenarios. The horizontal axis is lightest neutrino mass, i.e. $m_3$($m_1$) for inverted (normal) hierarchy. The allowed region for 3+1 flavors (3 flavors) shown in red (green) are calculated with best-fitted 3-flavor oscillation parameters from~\cite{PDG2014}. Mixing parameters of the additional sterile flavor is from~\cite{Kopp:2013vaa}. The dashed horizontal line indicates the targeted sensitivity for next-generation ton-scale neutrinoless double beta decay experiments.}   
\label{fig:double_beta}
\end{figure}

Neutrinoless double beta decay ($0\nu\beta\beta$) experiments are probing the Majorana nature of neutrinos and may allow determination of the effective neutrino mass from the nuclear transition rate.  
Light sterile neutrinos can contribute to the $0\nu\beta\beta$ decay rate ~\cite{Giunti:2015iyr} and thus alter the prediction of the effective neutrino Majorana mass $m_{\beta\beta}$ significantly as shown in Fig.~\ref{fig:double_beta}. 
In the 3+1 sterile neutrino model, the addition of sterile neutrino mass and mixing expands the allowed region for the inverted hierarchy and shifts the normal hierarchy space to higher values of $m_{\beta\beta}$. 
Next-generation, ton-scale $0\nu\beta\beta$ experiments aim to probe this effective neutrino mass down to 15~meV~\cite{NSACreport2015}. 
With this sensitivity, experiments will search the entire allowed parameter space of the inverted hierarchy under the assumption of 3 active neutrinos. 
Both understanding the mass hierarchy and the existence of sterile neutrinos will be important for interpreting the parameter space probed by $0\nu\beta\beta$ experiments. 

The existence of light sterile neutrinos would impact neutrino mass measurements from beta decay and cosmological experiments. 
An eV-scale sterile neutrino mass term contributes to the effective electron neutrino mass $m_{\beta}=\sum_{i} \sqrt{|U_{ei}^{2}m_{i}^{2}|}$ and alters the spectrum shape measured in beta decay experiments.
For this reason, sterile neutrinos may manifest itself in the Karlsruhe Tritium Neutrino (KATRIN) experiment, which can in turn probe the global parameter space in $\Delta$m$_{14}^2$ and $\sin^{2}2\theta_{14}$ that is complementary to reactor experiments~\cite{Katrin1eVPLB,Katrin1eVPRD}.  
$\Sigma  m_\nu$ measured in cosmological experiments such as Planck~\cite{Planck2015_XIII} is also sensitive to the existence of sterile neutrinos.
Based on current results, the observation of an eV-scale sterile neutrino in terrestrial experiments would be in tension with existing model-dependent cosmological limits and require additional model extensions such as sterile neutrino decay~\cite{SterileDecay}.

%----------------------------------------------------
% !TEX root = main.tex
\section{Worldwide Program to Search for Sterile Neutrinos: Complementarity and Discovery Reach}
\label{sec:intContext}

Multiple experiments have been proposed to test the eV-scale sterile neutrino interpretation of the anomalous results described in Sec.~\ref{sec:reactorAnomaly} and~\ref{sec:acceleratorAnomaly}.
These programs include observations of reactor \nuebar{}~\cite{Ryder:2015sma, Alekseev:2013dmu,Serebrov:2015ros, Pequignot:2015rta, Kim:2015qlu, Lane:2015alq}, as well as \nue{} and \nuebar{} from radioactive sources~\cite{Borexino:2013xxa}. 
All must operate at short baselines of several meters from a strong neutrino source. This leads to a unique set of challenges, including near-surface backgrounds, source related backgrounds, and facility constraints on possible baselines and/or target mass. 
To understand the physics reach of the proposed short-baseline efforts, we briefly summarize the range of parameters and their impact on the sensitivity of the experiments.

With the exception of the source experiments using pre-existing, large neutrino detectors, all of the proposed reactor experiments will be composed of liquid or solid scintillator targets with fiducialized masses on the ton-scale. 
Exposures, defined as the ``reactor power$\cdot$detector mass$\cdot$reactor live time'', are also similar for most next-generation detectors located $\mathcal{O}$(10
~meters) from HEU reactor cores, ranging from $\sim$40-157~MW$\cdot$ton$\cdot$year. 
PROSPECT Phase~I will have an exposure of 157~MW$\cdot$ton$\cdot$year. 
As such, the difference in sensitivity between different experiments depends on the baseline coverage, energy resolution, position resolution, and the signal-to-background (S:B) ratio. PROSPECT Phase~II proposed to add a second detector with mass of $\mathcal{O}$(10~tons) that increases the oscillation sensitivity beyond the general sensitivity of ton-scale, short-baseline experiments. The second detector will greatly increase the reach in baseline and the accessible range of the characteristic $L/E$ parameter for neutrino oscillations. 
 
All but one of the proposed experiments will have the capability to probe distances of $<$10~m from respective reactor cores maximizing both the \nuebar{} interaction rate and sensitivity to meter-scale oscillations.
However, only two of the eight reactor experiments, including PROSPECT, plan to vary the baseline with their ton-scale detector. In addition to improving the oscillation sensitivity, variable baseline coverage  provides an important handle on detector systematic checks and background studies. 

Good resolution in energy and position are required to make a measurement of the HEU spectrum, search for oscillations, and efficient background rejection. 
To test the observed spectral anomaly, an energy resolution of at least that obtained by the $\theta_{13}$ experiments, 8\% at 1~MeV, is ideal.
PROSPECT aims to achieve 4.5\%/$\sqrt{E}$ which is sufficient to resolve spectral features associated with significant groups of decay branches, and thus inform modeling efforts. With a position resolution of 15~cm PROSPECT will be able to resolve any oscillation effects and perform an efficient rejection of dominant backgrounds from cosmogenic showers.

The signal-to-background in the detectors varies depending on reactor site and overburden, passive shielding, and background rejection capabilities.  Some of the reactor experiments have done background studies at various reactor sites and in some cases coupled these with detailed  Monte Carlo simulation.  Sensitivities reported by each collaboration include the projected S:B, ranging from 1-3 at HEU sites and 5 at LEU sites. 
PROSPECT has done an extensive suite of background measurements \cite{Ashenfelter:2015tpm} and data-benchmarked simulations that indicate an expected S:B ratio of $>$1 as summarized in Table.~\ref{tab:params}. 

In addition to reactor experiments, short-baseline searches using intense beta decay sources can also search for oscillations in the disappearance channel. In particular, the CeSOX experiment plans to perform such a measurement with a 2-4~PBq $^{144}$Ce-$^{144}$Pr source at a baseline of 7.15~m from the center of the Borexino detector at Gran Sasso Laboratory~\cite{Borexino:2013xxa}. Antineutrinos will be detected via the inverse beta decay process and evidence of oscillations would be probed through measurement of either relative L/E distortion or deviations in the measured absolute rate and spectrum from predictions. %This technique is essentially background-free. 
Data will be taken for 1.5~years beginning in late 2016. This experiment will provide a sensitive probe of short-baseline oscillation, but source-related systematic estimates based on absolute activity measurement uncertainties and on neutrino spectrum shape factors will be challenging to produce. Run time and total statistics are limited for this experiment type by the 285 day half-live of the radioactive source.  

The Short Baseline Neutrino (SBN) program \cite{Antonello:2015lea}, a consortium consisting of three detectors at varied distances from the accelerator neutrino source at Fermilab, aims to perform a relative oscillation search to resolve the LSND and MiniBooNE anomalies. 
While reactor experiments pursue oscillations in \nuebar{} disappearance channel, SBN searches for hints of sterile neutrinos in \nue{} appearance channel in addition to providing a sensitive relative $\nu_{\mu}$ disappearance analysis.
The combined suite of experiments will help address all the relevant anomalies in the corresponding parameter space. 
Oscillation hints observed in one channel will suggest regions of focus for other channels. Conversely, a null oscillation result with significant confidence in all channels will conclusively rule out the existence of sterile neutrinos in the currently-suggested $|\Delta m^{2}|$$\sim$1~eV$^{2}$ parameter space.

%----------------------------------------------------
\section{PROSPECT Physics Program and Discovery Potential}
\label{sec:physics-program}
% !TEX root = main.tex

\subsection{Sensitivity to Short-Baseline Neutrino Oscillation}
\label{sec:oscSens}

\begin{figure}[tbp]
\centering
\includegraphics[ clip=true,width=0.48\textwidth]{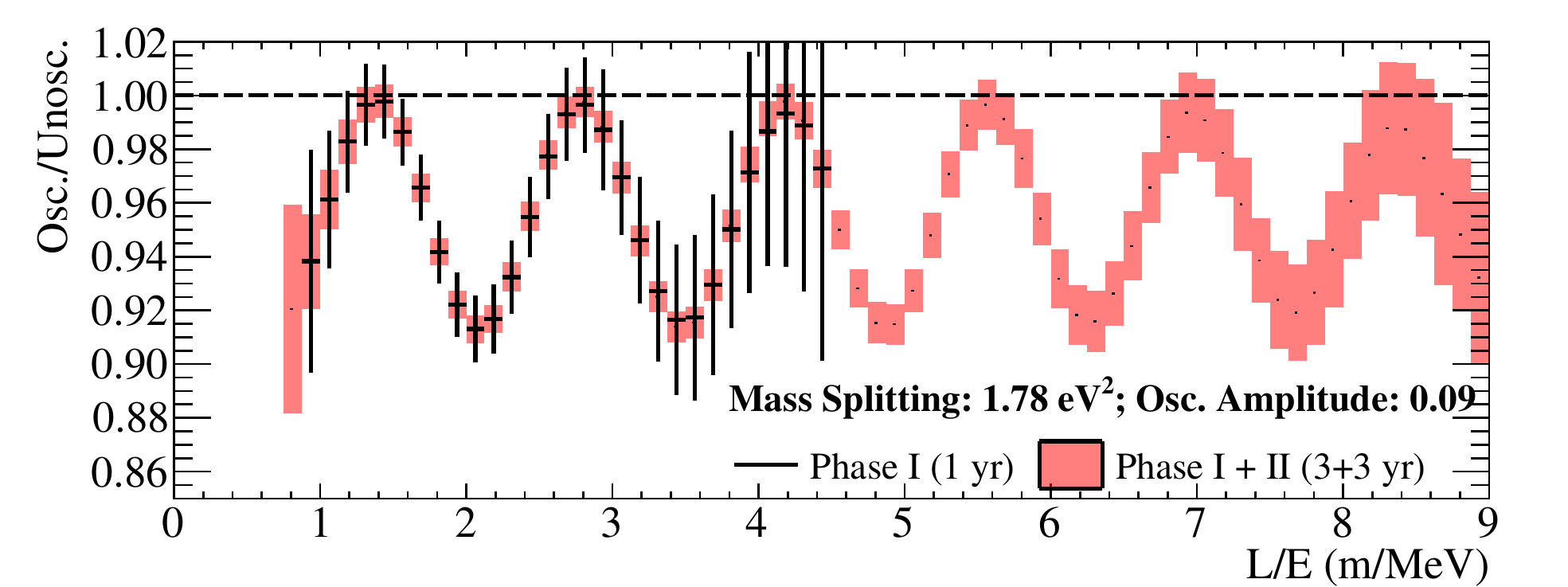}\\
  \vspace{-9pt}
   \caption{
   Asymmetry between oscillated and un-oscillated $L/E$ spectra after 1 year of Phase~I (black) and 3 years of both Phase~I+II for representative oscillation parameters.  
    }
   \label{fig:LoverE}
\end{figure}

PROSPECT will perform a sensitive search for light sterile neutrinos at the eV mass scale by probing signatures of neutrino oscillation through a relative comparison of the reactor flux and spectrum across a range of baselines (Fig.~\ref{fig:experimentalLayout}). %of $\le$11~m. 
The experiment has been designed (Sec.~\ref{sec:expStrat}) to provide  significant improvements in baseline coverage, event statistics, and energy resolution 
over previous short-baseline reactor oscillation measurements, thus providing coverage of oscillation parameter space that was previously inaccessible. 

To demonstrate the oscillation physics reach, we present sensitivity studies assuming the existence of one sterile neutrino in addition to the three known neutrino species, commonly referred to as the 3+1 model.
We note that PROSPECT's broad baseline range and in particular its extended reach in $L/E$ during Phase II will also provide sensitivity  to multiple sterile neutrinos~\cite{MultSBL}. 
The choice of many input parameters is informed by the \pspt{} R\&D program. This includes detailed information on the HFIR core, \adone{} performance and background estimates obtained from simulation studies and test detector operation, and deployment locations based on engineering engagement with the HFIR facility.  

\begin{figure}[h!]
\centering
\includegraphics[width=0.40\textwidth]{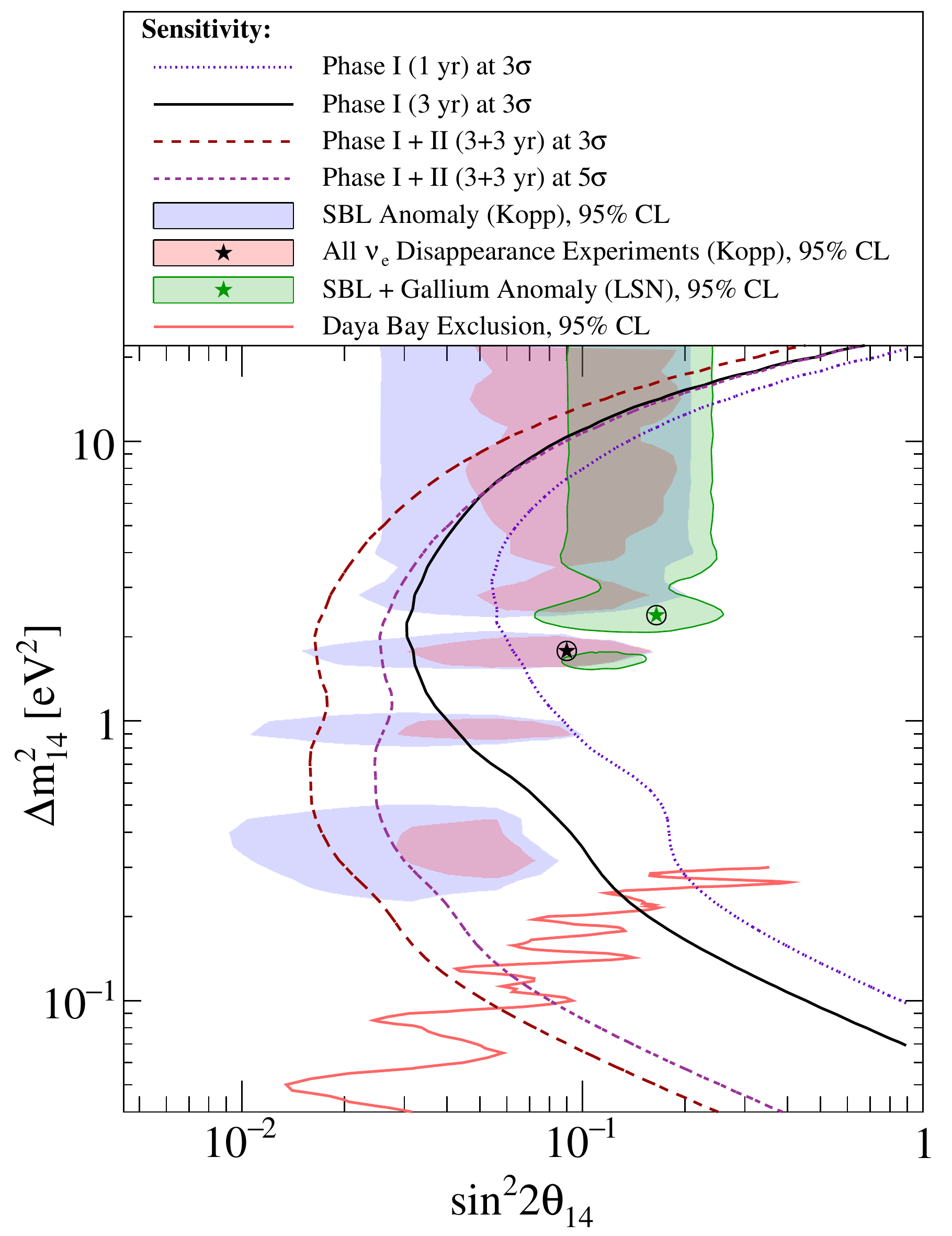} \\
\includegraphics[width=0.40\textwidth]{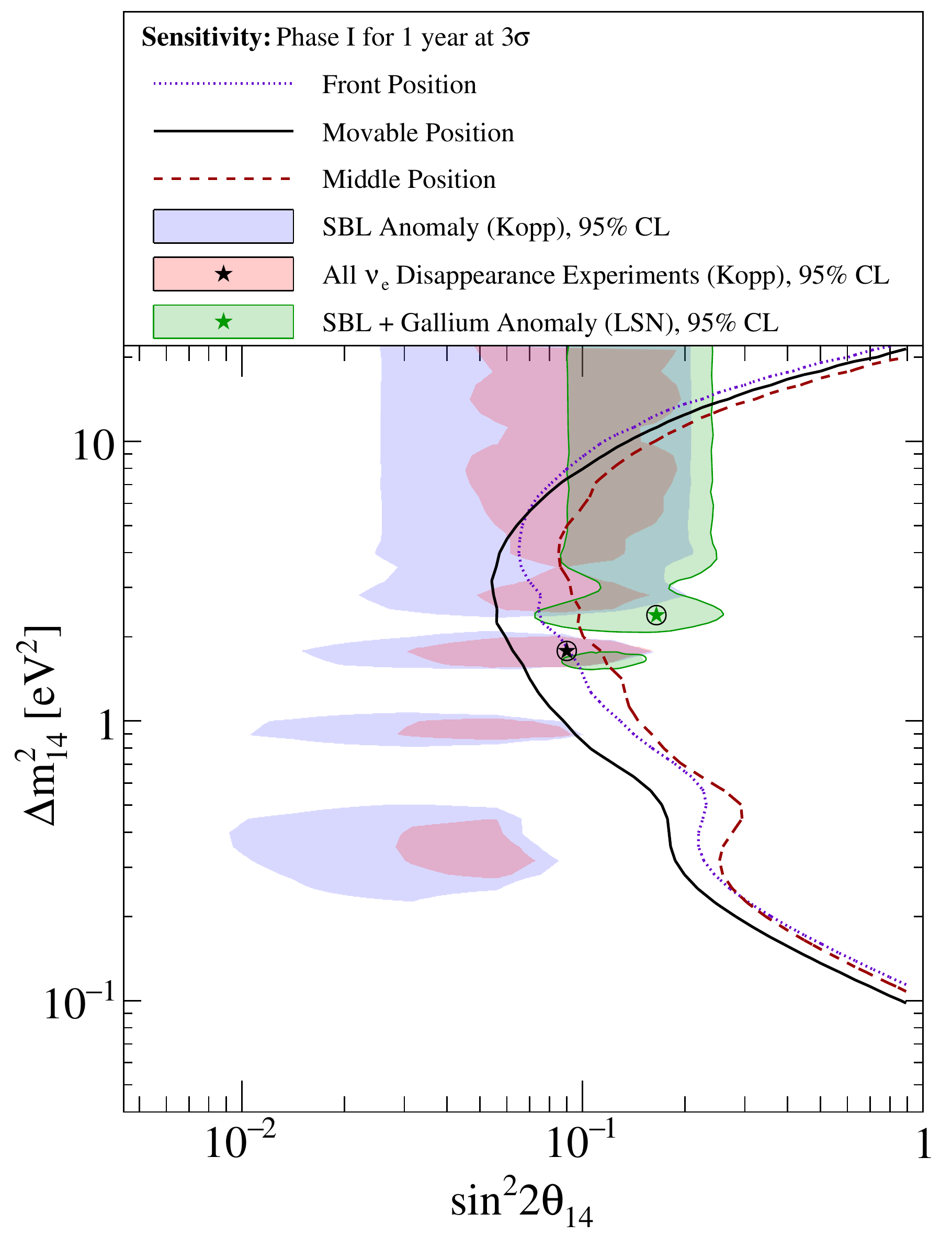} \\
   \caption{
	(Top) PROSPECT Phase~I and Phase~II sensitivities to a single sterile neutrino flavor. Phase~I probes the best-fit point at 4$\sigma$ after 1 year of data taking and has $>$3$\sigma$ reach of the favored parameter space after 3 years. The combined reach of Phase~I+II after 3+3 years of data taking yields a 5$\sigma$ coverage over the majority of the parameter space below $\Delta m^2_{14}$$\sim$10~eV$^2$. Daya Bay exclusion curve is from~\cite{DYBSterile}.
   (Bottom) Increase in oscillation sensitivity to sterile neutrinos during Phase~I by operating \adone{} at two positions instead of at the front or middle position only.
    }
   \label{fig:oscSens}
\end{figure}

PROSPECT will measure the relative \nuebar{} flux and spectrum as a function of reconstructed baseline and can directly map out the oscillation effect within the segmented detectors~\cite{VSBL}.  
This is shown in Fig.~\ref{fig:LoverE} for the single sterile neutrino hypothesis with parameters matching a global fit to $\nu_e$ disappearance data (Ref.~\cite{Kopp:2013vaa}, hereafter referred to as ``Kopp best-fit''). 
For this best-fit mass splitting, more than one full oscillation wavelength will be visible in PROSPECT Phase~I due to the wide baseline and energy range covered. 
Extension of \pspt{} to Phase~II accesses more oscillation cycles and adds statistical precision, thereby enhancing sensitivity.
It should be emphasized that the oscillation measurement in the PROSPECT \adone{} is a relative comparison between $L/E$ bins rather than between the flux measured in each \adone{} segment.
Because the relative range of baselines spanned by \adone{} is smaller than the \nuebar{} energy range, each segment contributes to the majority of $L/E$ bins and relative normalization plays a less important role in \pspt{} than near and far detector relative normalization does in the recent $\theta_{13}$ experiments.  
Furthermore, as \adone{} is moved, the relative contribution of each segment to a particular $L/E$ bin varies, reducing the effect of both correlated and uncorrelated systematic biases more efficiently than a single extended detector.

\begin{table}[tbp]
\begin{minipage}{0.45\textwidth}
\small
    \begin{tabular}{| l | l |} \hline
 {\bf Parameter} & {\bf Value}  \\
      \hline
{\bf Reactor} &  \\
Power & 85 MW \\
Shape & Cylinder \\
Size & 0.2~m $r$~$\times$~0.5~m $h$  \\
 Fuel & HEU \\
 Duty cycle & 41\% reactor-on \\ \hline
{\bf Antineutrino Detector 1 (AD-I)} &  \\
    Cross-section & 1.2$\times$1.45~m$^2$ \\
    Proton density & 5.5$\times$10$^{28}$~p/m$^3$ \\
    Total Target Mass & 2940~kg\\
    Fiducialized Target Mass&  1480~kg \\
    Baseline range & 4.4~m\\     
    Efficiency in  Fiducial Volume & 42\%\\
    Position resolution & 15~cm  \\
    Energy resolution & 4.5\%/$\sqrt{E}$ \\ 
    S:B Ratio & 3.1, 2.6, 1.8\\
    Closest distance & 6.9~m, 8.1~m, 9.4~m  \\\hline
{\bf Antineutrino Detector 2 (AD-II)} & \\ 
    Total Target Mass & $\sim$10~ton\\
     Fiducialized Target Mass & $\sim$70\% \\
    Baseline range & $\sim$4~m\\
     Efficiency in Fiducial Volume & 42\%\\
    Position resolution & 15~cm  \\
    Energy resolution & 4.5\%/$\sqrt{E}$ \\
    S:B ratio &  3.0 \\ 
    Closest distance & 15~m\\\hline
  {\bf Operational Exposure} &  \\
 Phase~I & 1, 3 years \\
 Phase~II & 3 years  \\ \hline
    \end{tabular}
  \caption{Nominal PROSPECT experimental parameters.  Phase~I consists of operating \adone{} for three years split between front, middle, and back positions. Phase~II adds \adtwo{} at a longer baseline and operates both detectors for three additional years.}
  \label{tab:params}
  \end{minipage} 
\end{table}

PROSPECT oscillation sensitivity is determined using a $\chi^2$ minimization~\cite{Ashenfelter:2013oaa}.
Systematic uncertainties are included by minimizing over nuisance parameters in addition to the new oscillation parameters ($\Delta m^2_{14}$, $\sin^22\theta_{14}$).
We take a conservative approach of allowing uncertainties for these parameters to vary broadly with little penalty in the fit. 
Relative uncertainties in normalizations and uncorrelated spectral variations between segments are assigned a 1\% uncertainty to match segment-to-segment differences observed in Monte Carlo simulations of the Phase~I \adone{}. 
A simulation-predicted background shape is used (see Fig.~\ref{fig:cosmic_IBD_spectrum_cuts}), and the signal-to-background ratio adjusted to account for the $1/r^2$ flux reduction at farther positions.

The sensitivity of PROSPECT to \nuebar{} oscillation after 1 and 3 calendar years (6 and 18 reactor cycles, respectively) is shown in Fig.~\ref{fig:oscSens}.  
In the first year of data taking, \adone{} will be operated equally at two positions separated by $\sim$1.5~m.  
The 3-year run will further increase baseline coverage with deployment at a third location separated by an additional $\sim$1.5~m from the front position. 
Within a single calendar year, \pspt{} can probe the best-fit of all current global analyses of \nue{} and \nuebar{} disappearance~\cite{Abazajian:2012ys,Kopp:2013vaa} at 4$\sigma$ confidence level.
Over 3 years of operation, \pspt{} Phase~I can discover oscillations as a sign of sterile neutrinos at $>$3$\sigma$ over the majority of suggested parameter space.
The sensitivity achieved with Phase~II is also shown: after 3 additional years of operation essentially all parameter space suggested by $\nu_e$ disappearance data below 10~eV$^2$ can be excluded. 

\begin{table}
   \centering
\small
   \begin{tabular}{|l|ccc|}
   \hline
	& \textbf{Decreased} & \textbf{Nominal} & \textbf{Increased} \\ \hline
	\textbf{Position} & \textit{Front only} & \textit{Movable} & \textit{Middle only}\\ 
	& 2.79 & 4.60 & 2.37 \\ \hline
	\textbf{Position} & \textit{10cm} & \textit{14.6cm} & \textit{20cm}\\ 
	\textbf{Resolution} & 4.69 & 4.60 & 4.46 \\\hline
	 \textbf{Efficiency} & \textit{32\%} & \textit{42\%} & \textit{52\%} \\ 
	 & 3.84 & 4.60 & 5.26 \\\hline
	 \textbf{Energy} & \textit{3\%} & \textit{4.5\%} & \textit{20\%} \\
	\textbf{Resolution} & 4.61 & 4.60 & 4.20 \\ \hline
	 \textbf{Background} & \textit{$\times$0.33} & \textit{--} & \textit{$\times$3} \\
	 \textbf{Suppression} & 3.92 & 4.60 & 5.00 \\  \hline
	 \textbf{Bin-to-Bin} & \textit{0.5\%} & \textit{1.0\%} & \textit{2.0\%} \\
	 \textbf{Uncertainty} & 4.69 & 4.60 & 4.30 \\ \hline
	 \textbf{Relative Segment} & \textit{0.5\%} & \textit{1.0\%} & \textit{2.0\%} \\
	 \textbf{Normalization} & 4.60 & 4.60 & 4.59 \\ \hline
	 \textbf{Detector} & \textit{10$\times$8} & \textit{12$\times$10} & \textit{14$\times$12} \\
	 \textbf{Size} & 3.23 & 4.60 & 6.02 \\ \hline
   \end{tabular}
   \caption{The effect of varying experimental parameters (\textit{italic}) on the confidence level in the unit of $\sigma$ with which oscillations at the Kopp best-fit point can be differentiated from the null hypothesis with one year of data-taking.}
   \label{tab:senseff}
\end{table}

The dependence of the sensitivity on experimental parameters is examined in Table~\ref{tab:senseff}. These results clearly validate the design focus on background rejection and maximizing target mass, while also highlighting the value of covering the widest possible baseline range via movement of \adone{}.
The increase in sensitivity afforded by the expanded $L/E$ coverage gained though \adone{} movement is further illustrated in the bottom panel of Fig.~\ref{fig:oscSens}. 
Although the signal decreases as the inverse of $r^2$, the gain from $L/E$ coverage surpasses the loss due to reduced signal when the detector is operated equally at two positions. 
It must be noted that for the sensitivity calculation shown this gain is purely from the improved $L/E$ coverage. Moving the detector also gives a better control of correlated and uncorrelated systematic biases, which can be expected to further increase the sensitivity. 
The ultimate choice of positions will be guided by the demonstrated S:B at various baselines. 

\subsection{Precision Measurement of the Reactor \nuebar{} Spectrum }

PROSPECT will measure the energy spectrum of \nuebar{} emitted by an HEU reactor with a precision that exceeds previous experiments and current model predictions. 
Between 2--6~MeV, Phase~I will achieve an average statistical precision better than 1.5\% and systematic precision better than 2\%.
The target energy resolution, $4.5\%/\sqrt{E}$, will be greater than any previous reactor experiment and will allow for the detection of fine structure in the antineutrino spectrum.
In contrast to LEU reactors where the fission fractions change as plutonium isotopes are produced in the core, the simpler HEU system allows for a more accurate evaluation of reactor evolution and flux prediction models.  

\begin{figure}[tbp]
\centering
    \includegraphics[width=0.40\textwidth]{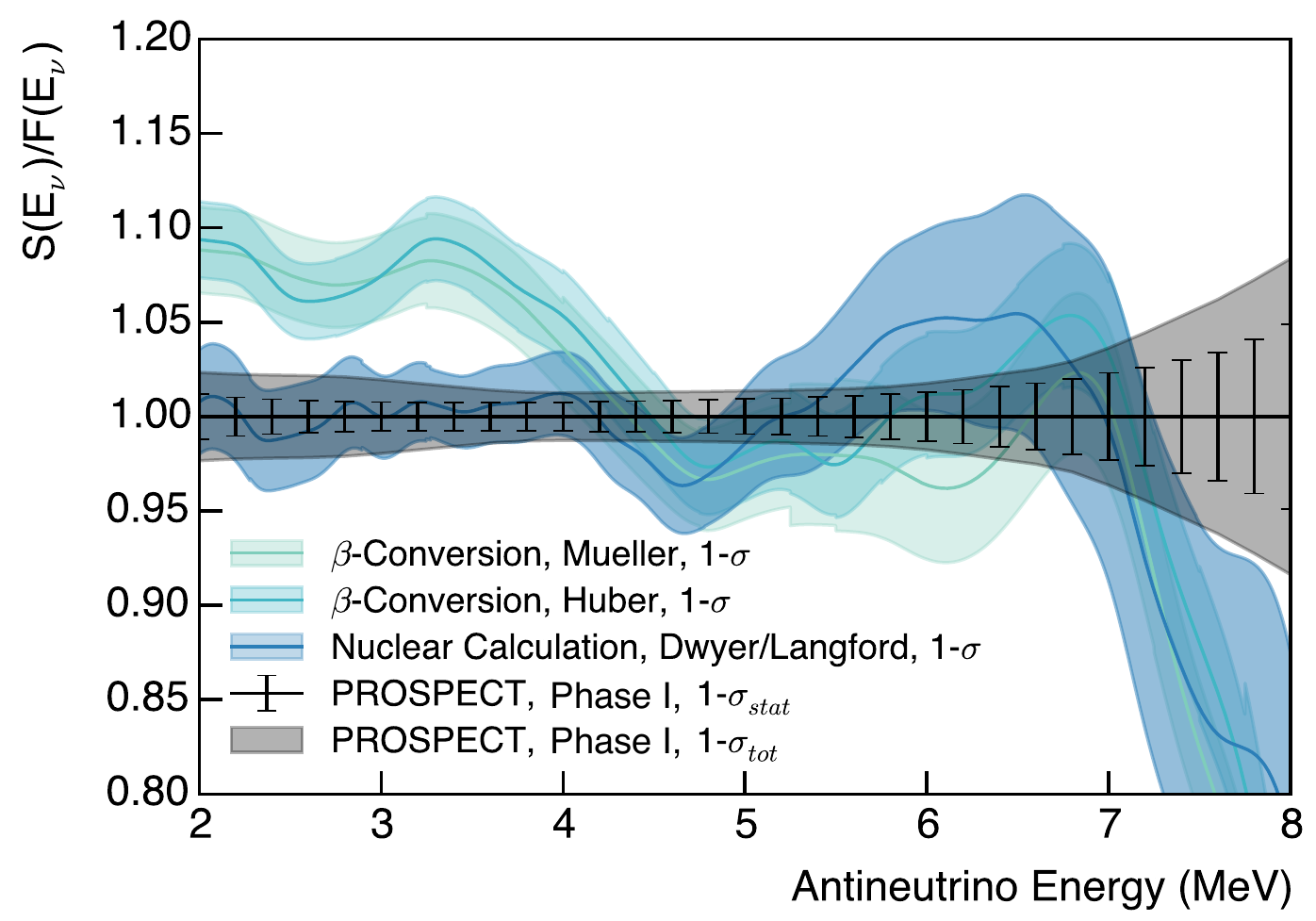}
	\includegraphics[width=0.40\textwidth]{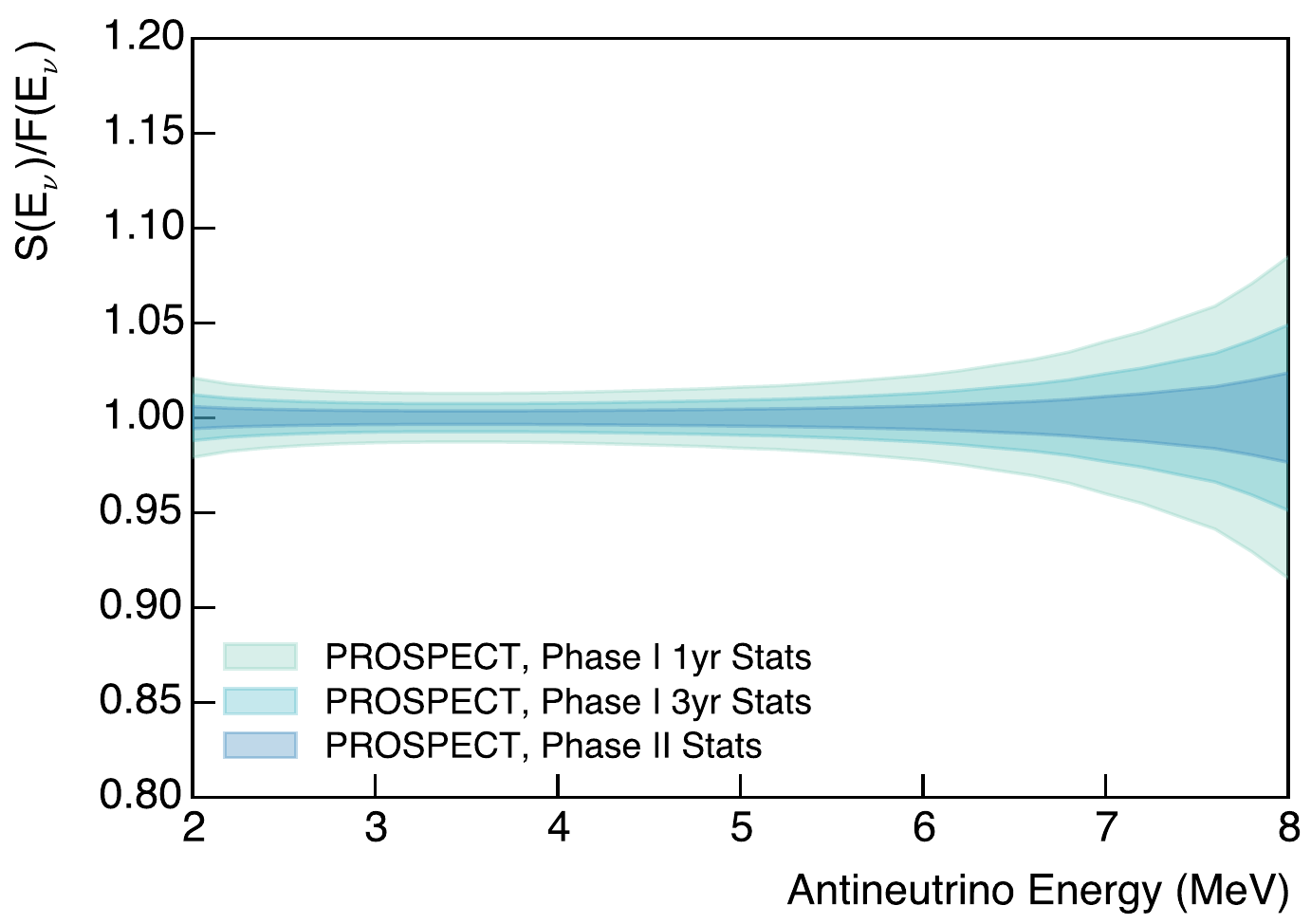}
   \caption{
   (Top) Three models of the $^{235}$U \nuebar{} energy spectrum relative to a smooth approximation.
The $1\sigma$ error band of the Phase~I measurement including subtraction of predicted background (error bars) and systematic uncertainties (gray band) are shown for comparison. An energy resolution of 4.5\%$/\sqrt{E}$  has been applied to highlight accessible features.
(Bottom) Evolution of statistical error bands for 200~keV bins from Phase~I to Phase~II.}
\label{fig:specModelComparison}
\end{figure}

Fig.~\ref{fig:specModelComparison} shows the differences between three current models: two based on the $\beta^-$-conversion method, and one based on {\em ab-initio} calculation.
To highlight the shape differences between models, they are shown in ratio to a smooth approximation $F(E)$\footnote{$F(E_{\overline{\nu}})={\rm exp}(\sum_i\alpha_iE_{\overline{\nu}}^{i-1})$, with $\bm{\alpha}=\{$1.418, -0.6078, 8.955$\times$10$^{-3}$, -6.690$\times$10$^{-3}$, 6.933$\times$10$^{-5}\}$}.  
The PROSPECT Phase~I statistical precision is shown as the black error bars.  
PROSPECT will be able to discriminate between these models and directly measure the spectrum more precisely than any of the predictions.  
In addition, this measurement can be combined with those underway at LEU reactors to extract the non-$^{235}$U contribution to the spectrum.  
Since current LEU measurements, and that of HEU which we propose, are expected to have percent-level precision, differences should be prominent and provide another route to evaluate and refine reactor models.

The segmented \adone{} detector is designed to enhance the spectral measurement through careful optimization of detector uniformity and construction techniques.
The use of low-mass reflector panels, described in more detail in Sec.~\ref{sec:sigResponse}, minimizes the non-scintillating volume that could bias the detected energy spectrum.
Multiple fiducialization schemes are being studied to determine the optimum volume selection that maximizes detection efficiency of positron annihilation gammas.

\adtwo{} is designed to achieve at least equal statistical power to that of \adone{}, even at a longer baseline.
A larger target mass and improved cosmogenic shielding increase the IBD detection rate without decreased signal-to-background ratio.
Both antineutrino detectors \adone{} and \adtwo{} are comprised of identical segments, ensuring that systematic uncertainties will be consistent.
Thus, all development and characterization of \adone{} will directly apply to \adtwo{}, simplifying the combined analysis during Phase~II.

Fig.~\ref{fig:specModelComparison} shows a comparison of the statistical error bands, assuming 200~keV binning, for each experimental phase.
With Phase~II, PROSPECT will achieve an average of 1.0\% statistical uncertainty throughout the reactor antineutrino energy range.
With the combined phases, PROSPECT will have major statistical power to resolve and probe the spectral anomaly region and directly measure fine structure in the antineutrino spectrum.

\subsection{Reactor Monitoring and Characterization}

\begin{figure}[tbp]
\centering
    \includegraphics[trim={0cm 0cm 0cm 0cm},clip,width=0.40\textwidth]{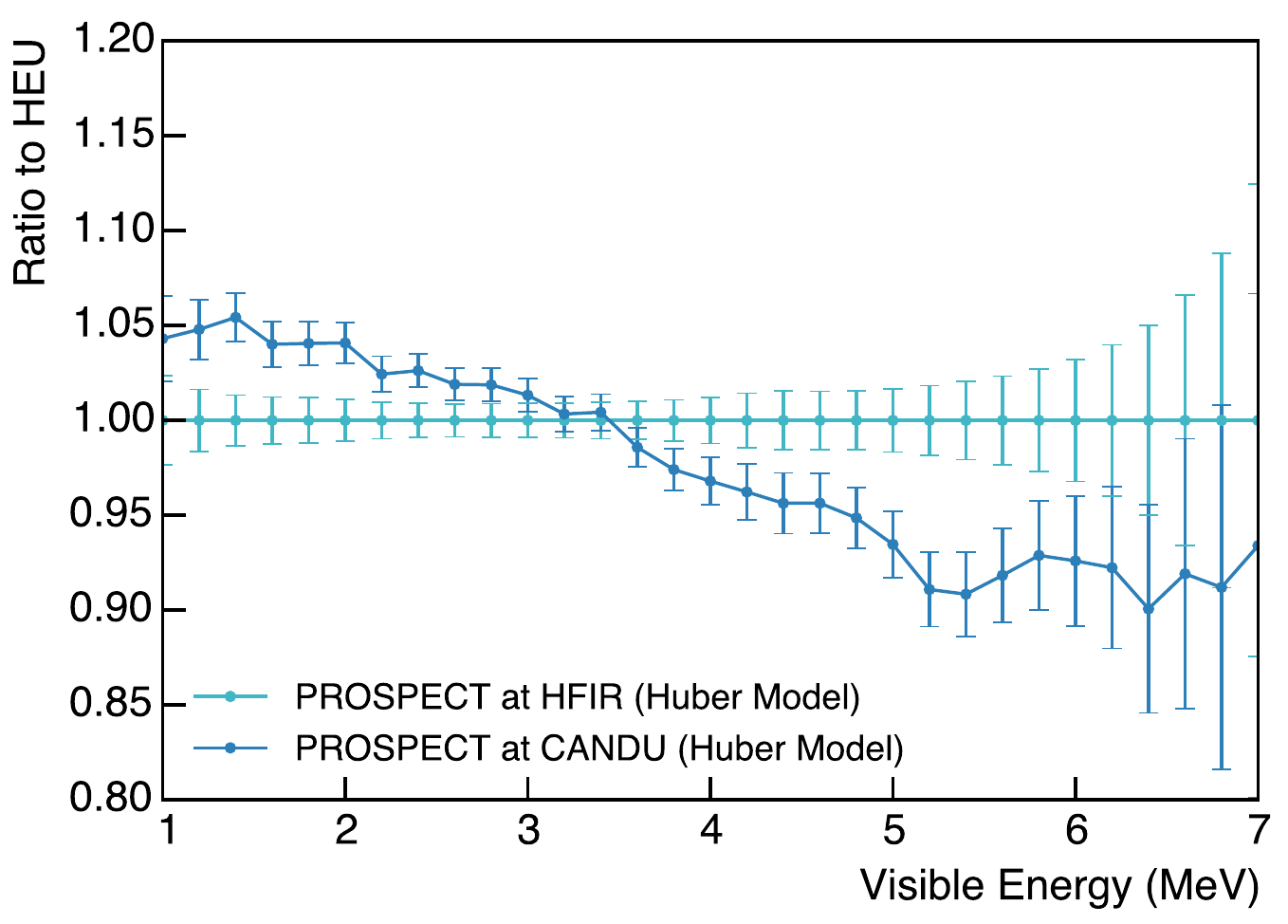}
   \caption{Ratio  of 3 year simulated measurements at HFIR and a CANDU reactor, assuming identical statistics and backgrounds.
Only statistical uncertainties are shown, as systematics will cancel. 
This \textit{time independent} difference in spectra is due to differing fuel mixes.}
\label{fig:specCANDUComparison}
\end{figure}
Phase~I of the \pspt{} experiment will develop technology, produce scientific results, and construct a detector that could find utility in other areas. 

While the highest priority for future spectrum measurements are HEU-fueled reactors, examination of other reactor types is also valuable. 
Measurement of a LEU reactor spectrum with the superior energy resolution of the PROSPECT \adone{} would supplement current statistically precise measurements ~\cite{An:2015nua,Abe:2014bwa,Kim:2014rfa}, improve the knowledge of \nuebar{} spectra from fission of \isot{238}{U}, \isot{239}{Pu} and \isot{241}{Pu}, and reduce systematic uncertainties in the comparison of LEU and HEU \nuebar{} spectra through use of a common detector for both measurements.  
Measurement of a CANDU reactor~\cite{CANDU, Torgerson20061565}, in which frequent refueling maintains a static fuel mixture of $^{235}$U, $^{238}$U,  $^{239}$Pu, and $^{241}$Pu, would further improve the determination of each spectral component. 
Reactors with different core neutron spectra, which populate different fission daughter distributions, should also be considered. 
\pspt{} has contacts at several candidate reactors for such follow-up measurements after its run at HFIR.

Compact  \nuebar{} detector development is also of interest for reactor monitoring applications~\cite{Bernstein:2001cz,ApplNucl, NUDAR:2013}.  
\pspt{} will advance these efforts through development of background rejection techniques, demonstration of near-surface \nuebar{} detection, and precision \nuebar{} spectrum measurement in compact detectors. These capabilities are of interest to potential end users of reactor monitoring technology~\cite{IAEA2009,IAEA2012}. Successful demonstration of reactor \nuebar{} flux and spectrum measurements near-surface would open a range of possibilities for deployment at various baselines that would otherwise be inaccessible.

% !TEX root = main.tex

\section{PROSPECT Experimental Strategy}
\label{sec:expStrat}
%\graphicspath{ {./figs/}

PROSPECT is conceived as a phased experiment involving two independent detectors at multiple baselines. 
Phase~I consists of a movable antineutrino detector, \adone{}, with a 3-ton active target mass at baselines ranging from 7--12m from the reactor core. 
Phase~II of the experiment involves an additional second detector, \adtwo{}, of $\sim$10-tons deployed at 15--19~m from the core. 
High-impact first physics results can be produced after 1 year of data taking in Phase~I. The physics goals of Phase~I can be completed with 3 years of data taking of \adone{}. 
The ultimate physics reach of PROSPECT is obtained after an additional 3 years of simultaneous operation of \adone{} and \adtwo{}.

\begin{figure*}[t]%{R}%{0.65\textwidth}
  %\vspace{-10pt}
    \centering
   \includegraphics[trim={0.0cm 0.0cm 0cm 0cm},clip,width=0.7\textwidth]{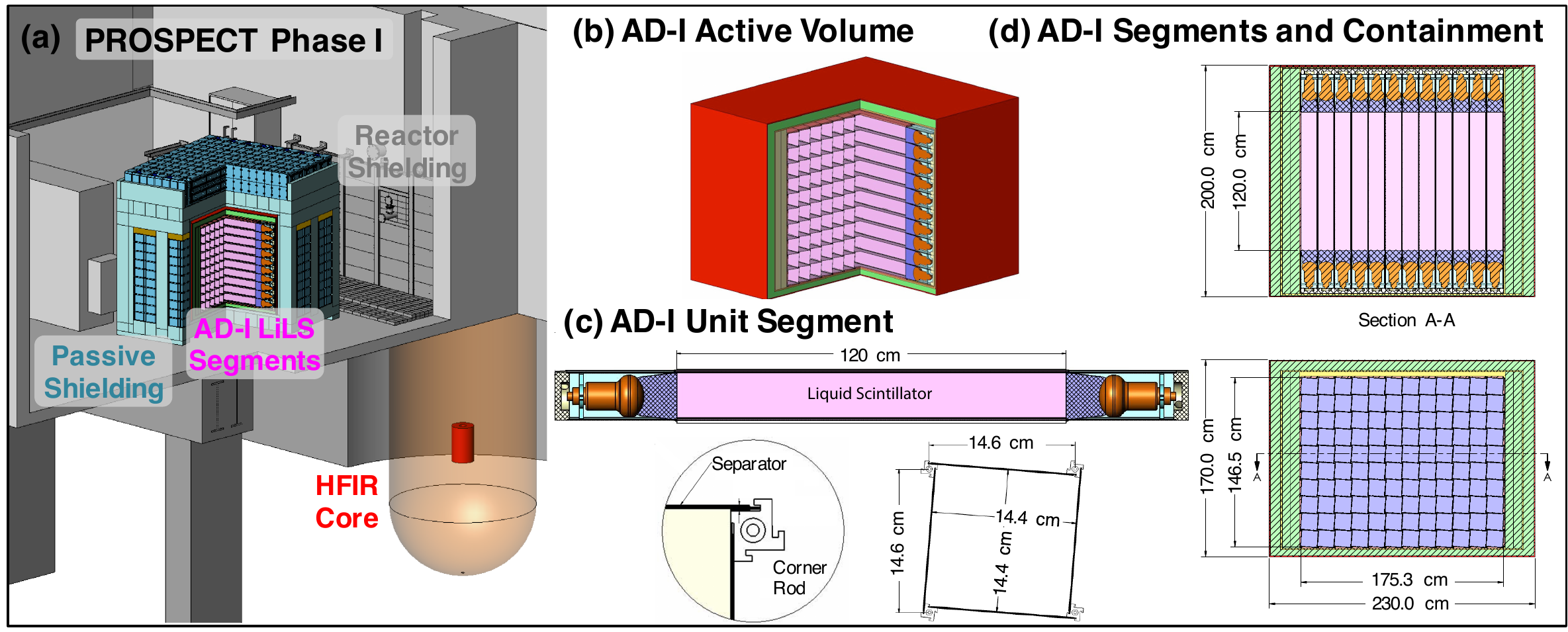} 
  %\vspace{-15pt}
\caption{(a) Model of \adone{} for Phase~I at HFIR. (b) A cutaway diagram of the \adone{}. (c) The unit segment structure. (d) Inner and outer dimensions of the \adone{}.} 
   \label{fig:phaseI-design}
  %\vspace{-10pt}
\end{figure*}

\subsection{Reactor Site}

After a thorough study of three possible US research reactor sites, all of which could support the experimental goals, \HFIR{} at \ORNL{} was selected for Phase~I of \pspt{}. 
Facility parameters, including the size and power of the compact HEU fueled core, and operational duty cycle are given in Table~\ref{tab:params}.
Deployment locations for both \adone{} and \adtwo{}, shown in Fig.~\ref{fig:experimentalLayout}, have excellent access and controllable reactor-correlated and cosmogenic background levels. 
Through extensive engagement with HFIR, it has been established that \adone{} and the associated passive shielding design meets all space, floor-loading, and safety requirements and would permit a $\sim$3~m range of horizontal movement.
%as seen in (Fig.~\ref{fig:phaseIExperimentalConcept}).

Extensive background measurements at \adone{} locations have identified specific ``hot spots'' that can be reduced with localized shielding (Sec.~\ref{sec:bkgMeasure}). 
There is no significant reactor-correlated background at the \adtwo{} location.
Prototype detector and shielding tests show that reactor-produced $\gamma$-ray and neutron backgrounds can be suppressed to insignificant levels with appropriate shielding (Sec.~\ref{sec:p20onsite}).
Remaining time-correlated backgrounds are dominated by cosmogenic fast neutrons since the detector sites have minimal overburden. 
Shielding for \adone{} will therefore have fixed lead walls to control reactor backgrounds and shielding that moves with the detector to reduce cosmogenic backgrounds.

\subsection{Antineutrino Detector Design and Performance}
\label{sec:detectorDesign}

The PROSPECT antineutrino detector (\adone{}) for Phase-I consists of a single volume of $^6$Li-loaded liquid scintillator (LiLS) segmented by low-mass high-reflectivity optical separators. 
The \adone{} detects reactor \nuebar{} via the inverse beta decay (IBD) reaction $\overline{\nu}_{e} \rm{+p} \rightarrow e^+\rm{+n}$. 
The positron carries most of the \nuebar{} energy and makes a prompt energy signal in the LS. The neutron thermalizes before capture on $^6$Li or hydrogen, producing a delayed signal $\sim$40~$\mu$s later. $^6$Li doping 
was chosen as its decay products produce easily recognized, localized energy depositions in pulse shape discrimination-capable LS. 
This time-correlated signature of a gamma-like prompt signal and a neutron-capture-like delayed signal is extremely effective at nearly eliminating randomly time-coincident (accidental) backgrounds. 

The optical separators divide the total active volume ($\sim$3000~l) into 120 individual segments (Fig.~\ref{fig:phaseI-design}) providing baseline and event topology information independent of light transport and timing.   
Each segment shares optical separator panels and hollow support rods with its nearest neighbors and is read out at both ends by photomultiplier tubes (PMTs).   
Space constraints have largely determined the designed segment length, while cross-section dimensions are constrained by the physical dimensions of the PMTs and their housing assemblies.
%The double-ended readout improves light collection efficiency and PSD capability and enables position reconstruction along cell axis using the relative charge and time of the two PMT signals.
To maintain LiLS compatibility, the PMT and its voltage divider are housed inside a polycarbonate module with a light guide for optical coupling. 
%Signal and HV cables  exit through seal plugs on the other end. 
Modules are bolted together (10 high $\times$ 12 long) to form a support structure for the optical separator array. 
The separator panels and corner rods are designed to minimize inactive material and amount to 1.8\% of the total target mass. This is significantly less 
than earlier experiments such as Bugey~3 (15.5\% inactive mass) \cite{Achkar1995503}.
A carefully selected subset of the support rods house either optical fibers or tubes containing movable radioactive sources to calibrate segment energy response and timing. 
Cables, fibers, and calibration tubes are routed to the top surface, and this inner detector structure is inserted into a sealed acrylic single-volume LiLS containment vessel that isolates the inner detector from outside moisture and oxygen.
%Feedthroughs on  the acrylic lid and an O-ring seal on the perimeter isolate the inner detector from outside moisture and oxygen.  
%The acrylic vessel is filled with LiLS, including the active volume between optical separators and 
All space between PMT modules is filled with LiLS. 
The inactive LiLS not viewed by PMTs acts as additional passive shielding and totals $\sim$300~l.  
%The inner acrylic vessel is contained within an aluminum vessel that defines the \adone{} outer boundary.  
%The \adone{} will be positioned using a HFIR-owned Leica optical coordinate measuring machine to a precision of 0.5~cm. 
%The following sections describe the components of the \adone{} in more detail.

The \pspt{} LiLS has been developed over several years to exhibit the light yield and pulse shape discrimination (PSD)  required for the experiments physics (energy resolution) and background rejection (PSD) goals (Sec.~\ref{sec:lils}). 
Enriched $^6$Li, fluors (PPO) and wavelength shifter (bis-MSB) are added to a commercial scintillator base (EJ-309, Eljen Technologies~\cite{ej309}). 
Prototype studies in a 20~l, 1~m-long, test detector have demonstrated that a detected light yield of 6500 photons per MeV and a bulk attenuation length of 4 m are achievable, leading to an energy resolution of better than 4.5\%/$\sqrt{E}$.
The excellent PSD performance demonstrated in Sec.~\ref{sec:p20offsite} enables cuts preserving 99.9\% of the (n,Li) signal while rejecting the same fraction of $\gamma$-ray events.

Detailed simulations based on measured backgrounds at  HFIR and performance of prototypes indicate that PROSPECT can achieve and exceed the required signal-to-background performance of 1:1 (Fig.~\ref{fig:cosmic_IBD_spectrum_cuts}). Further details are given in Sec.~\ref{sec:bkg}.

\begin{figure*}[t]%{R}%{0.65\textwidth}
%  \vspace{-05pt}
    \centering
   \includegraphics[width=0.95\textwidth]{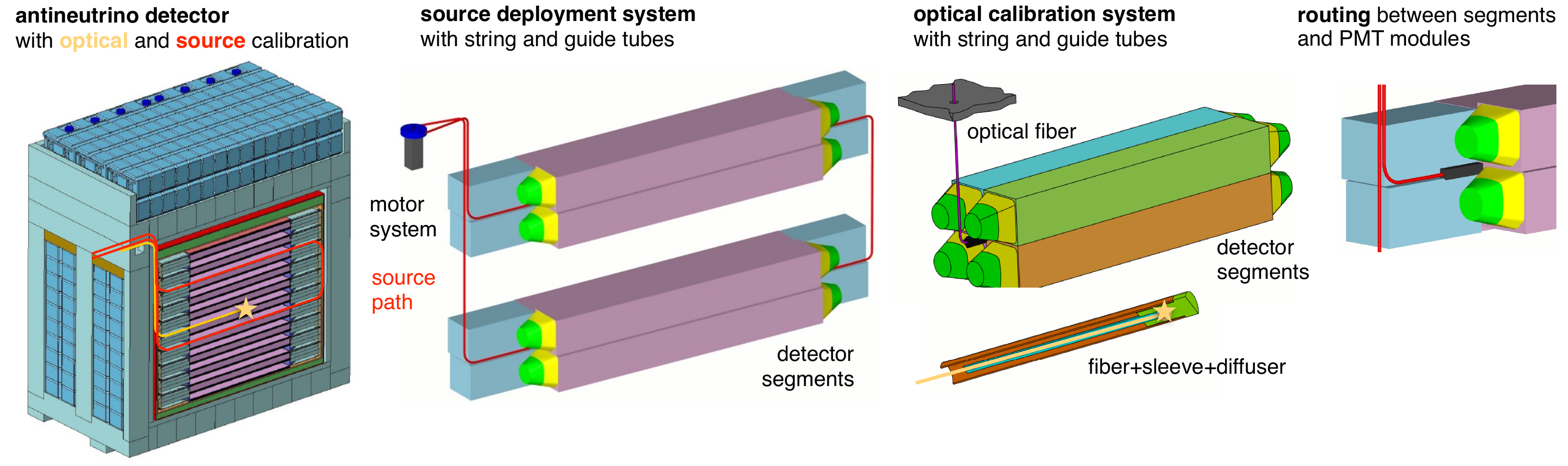} 
%  \includegraphics[trim=0.5cm 1.0cm 0.5cm 12.cm, clip=true, width=0.9\textwidth]{figs/Combined_source.pdf} 
 %\includegraphics[trim=0.4cm 0.3cm 0.4cm 0.1cm, clip=true, width=0.65\textwidth]{figs/gammacalib-082315.pdf} 
% \vspace{-7pt}
\caption{The \adone{} source deployment and optical calibration systems. Radioactive sources and an optical system will be deployed between detector segments.} 
   \label{fig:cal_system}
 % \vspace{-15pt}
\end{figure*}

%------------------------------------------------------------
%\subsection{Phase II Detector Design and Performance}
%\label{sec:detectorDesignPhaseII}

For Phase~II, a second antineutrino detector, \adtwo{}, would be installed just outside the HFIR reactor building covering baselines from 15--19~m.  
The detector features an increased volume of $\mathcal{O}$(10~tons), while maintaining the same segmentation as \adone{}.  
By using identical segment geometries, systematic uncertainties related to relative detector efficiency can be better controlled and confidence in the projected background rejection is increased.  
The active detector would be shielded by $\sim$0.75~m of steel and 1~meter of polyetheylene or water (nearly 5~m of water equivalent mass) to reduce cosmogenic backgrounds. 
Simulations predict a signal-to-background ratio of about 3.0, comparable to the closest \adone{} position from the reactor core.

\subsection{Calibration Strategy}
\label{sec:calib}

%\begin{figure}
%%\begin{wrapfigure}{R}{0.45\textwidth}
%% \vspace{-17pt}
%\centering
%\includegraphics[clip=true, trim=5mm 10mm 15mm 15mm,width=0.6\textwidth]{figs/08_21_2015_Calibration.pdf}\hfil
%%  \vspace{-10pt}
% \caption{\textbf{Top:} Stability calibration of  PROSPECT-20 detector at HFIR with deployed $^{60}$Co, cosmic $\mu$, neutron capture on $^{6}$Li and in-situ $^{40}$K sources over two months.  \textbf{Bottom:} Simulated PE spectra for three calibration sources, AmBe 4.4~MeV $\gamma$-ray (dark blue), neutron capture on H (red) and $^{137}$Cs (light blue) for \textbf{(Left)} the full \adone{} and \textbf{(Right)} an individual cell (by requiring no deposition in others cells). Full absorption peaks are available in both cases.}
%	%Background is primarily from cosmic fast neutrons.}
%%  \vspace{-15pt}
%\label{fig:cal_gamma_ray}
%%\end{wrapfigure}
%\end{figure}
%The \pspt{} calibration program is of central importance for achieving the physics goals, particularly for normalizing cell response and for establishing the \adone{} energy scale. 
%\pspt{} will adapt techniques proven in recent reactor \nuebar{} experiments to the segmented \adone{}, while also developing techniques to fully exploit new capabilities enabled by that geometry.  

The segmented \adone{} design incorporating hollow support rods will allow extensive access to the full \adone{} volume for routine calibration using optical fibers or retractable radioactive sources (Fig.~\ref{fig:cal_system}). 
LiLS light transmission, PMT gain, and PMT timing will be calibrated and monitored with a stabilized pulsed laser source via optical fibers, with each fiber illuminating 4 segments at their midpoint.  
Encapsulated $\gamma$-ray and neutron sources on tensioned string loops will be periodically deployed at multiple locations within the \adone{} via Teflon guide tubes in the support rods.

Fitting the deposited energy spectra of radioactive sources will allow the absolute positron energy scale, including scintillator non-linearity, to be calibrated to a few percent or better. 
These sources can also be used to ensure that small expected differences in positron energy scale between fiducial segments can be characterized to the percent level and corrected for in PROSPECT's sterile oscillation analysis.
Neutrons from encapsulated AmBe sources will allow calibration of PSD and determination of neutron detection efficiencies in each segment.  
Radioactive and cosmogenic backgrounds will be used to monitor and calibrate the detector response between source deployments, following the example of PROSPECT-20 which used \isot{40}{K}, neutron capture on $^6$Li, and through-going muons.
Finally, the possibility of spiking the scintillator with ${\cal O}(10^{-13}\ {\rm g})$ of \isot{227}{Ac} to exploit the double-$\alpha$ cascade from \isot{219}{Rn}$\to$\isot{215}{Po}$\to$\isot{211}{Pb} is being examined. 
This will allow a measurement of the uniformity per segment to 1\% and enable a relative LiLS mass measurement. 
Further R\&D is needed to ensure that dissolution and uniform distribution is possible without introducing unwanted backgrounds.

% !TEX root = main.tex

\subsection{Research and Development Status}
\label{sec:r-d-status}

The PROSPECT collaboration has conducted a vigorous R\&D program since 2013~\cite{Ashenfelter:2013oaa} and is exceptionally well prepared to perform the physics measurements described in Sec.~\ref{sec:physics-program}. 
Here we describe the central elements of that R\&D program including logistics and background studies at multiple reactors, detector design, liquid scintillator development, prototype operation, and simulation development and validation. 
Collectively, these efforts  demonstrate that the PROSPECT \adone{} design for Phase~I can be installed in a research reactor facility,  meet the necessary performance requirements, and that both reactor-generated and cosmogenic backgrounds can be controlled. 

\subsubsection{Reactor Facility Background Measurements}
\label{sec:bkgMeasure}

\begin{figure}
\centering
\includegraphics[trim={0cm 2cm 2cm 1cm},clip,width=0.40\textwidth]{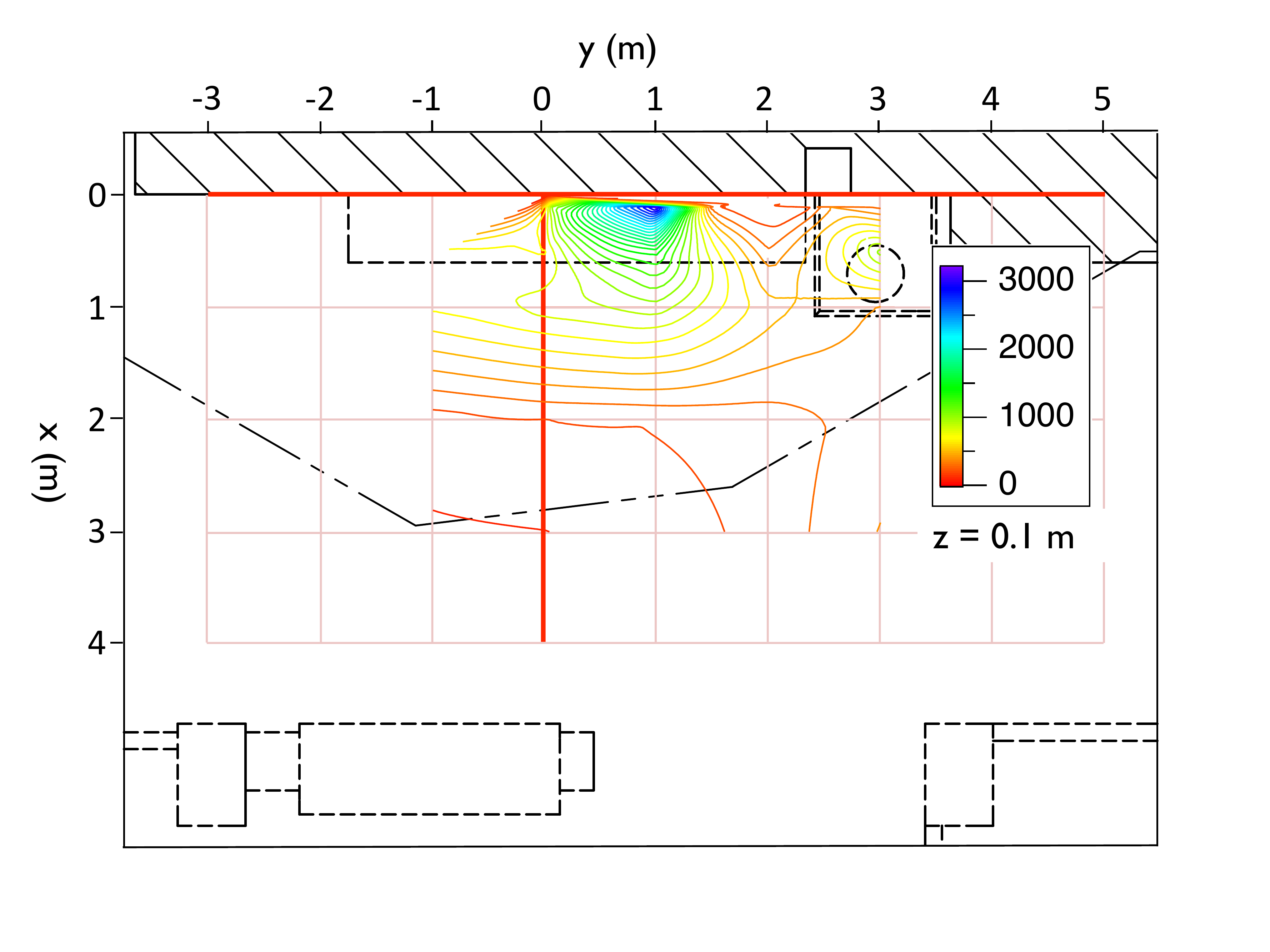}\\
\includegraphics[width=0.40\textwidth]{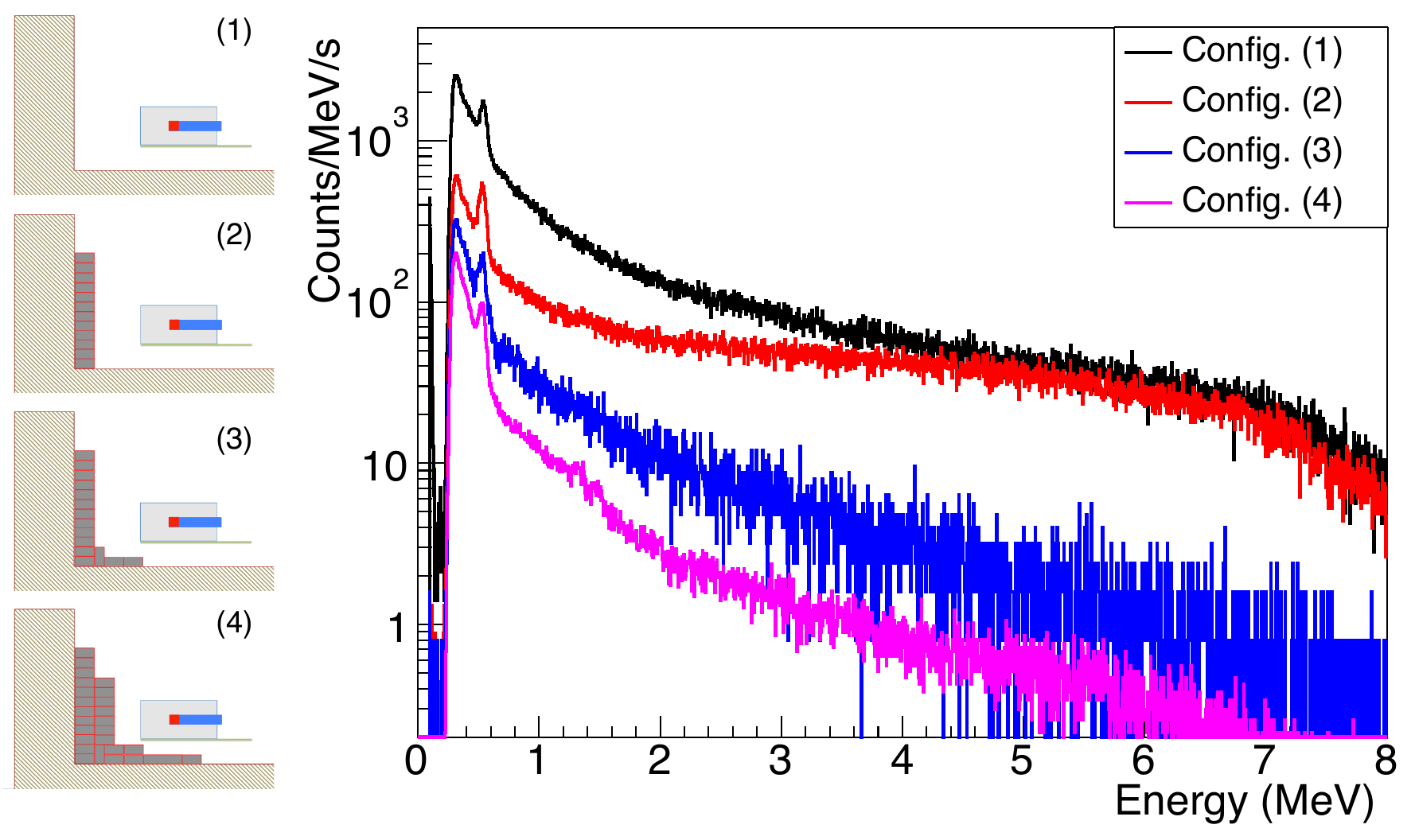}
\caption{ (Top) $\gamma$-ray count rates (s$^{-1}$) for an unshielded 2" NaI(Tl) detector with  HFIR at power. The core is at $(x,y,z) = (-4.06,0,-3.85)$. Strong spatial variation is caused by a plugged beam tube located at $(x,y,z) = (0,1,0)$. 
(Bottom) Energy spectra for a horizontally collimated 2" NaI(Tl) detector for varying configurations of a  lead wall, with HFIR at power. The shield is centered at the location of highest $\g$-ray intensity, $(x,y,z) = (0,1,0)$. The background rate is significantly reduced, even at high energies.}
  \label{fig:rxBkgMeas}  
\end{figure}

To obtain the broadest sensitivity to the possible existence of additional neutrino states~\cite{VSBL} and to maximize the event rate for a precision \nuebar{} energy spectrum measurement, Phase~I PROSPECT must be placed as close to a reactor core as is practical. 
In such a location, $\gamma$-rays and neutrons produced by reactor operation cannot be neglected, and indeed would be the dominant background source without careful attention to shielding.  
Furthermore, most facilities have minimal overburden, thus cosmogenic backgrounds are significant relative to the expected signal rate.  
In particular, fast neutrons from air showers can yield backgrounds that are challenging to shield, either with passive or active approaches.

The PROSPECT collaboration has conducted a careful assessment of natural and reactor generated background radiation that is reported in~\cite{Ashenfelter:2015tpm}. 
These measurements included high and moderate resolution $\gamma$-ray spectroscopy, fast and thermal neutron flux, muon flux, and fast neutron spectroscopy.  
Reactor facilities exhibit significant spatial variation in both $\g$-ray and neutron backgrounds due to irregular shielding, localized shielding leakage paths, or piping carrying activated materials (Fig.~\ref{fig:rxBkgMeas}), thus site-specific characterization of background is essential to optimize a shielding design. 
Localized shielding applied to compact background sources can be a cost- and weight-efficient approach to reducing background.  
Such an approach has been demonstrated quite successfully at the \HFIR{} site (Fig.~\ref{fig:rxBkgMeas}).  
The measurements described in Section~\ref{sec:p20onsite} show that targeted shielding, in addition to a carefully designed shielding package, can yield excellent control of reactor related backgrounds.
 
The flux and spectrum of cosmogenic fast neutrons observed within the minimal overburden provided by the HFIR building is essentially unaltered compared to standard reference measurements (e.g.~\cite{Gordon2004,Kowatari2005}), which can therefore be used as source terms in simulation studies. 
The results of these studies have been integral to the design of the Phase~I PROSPECT \adone{} and have been validated using a series of prototype detectors as described in Section~\ref{sec:p20offsite}. 
The detailed understanding of fast neutron related backgrounds afforded by this work has enabled the development of a series of effective analysis cuts that yield an expected S:B of better than 1:1 as described in Section~\ref{sec:bkg}.

\subsubsection{Liquid Scintillator Development}
\label{sec:lils}

Liquid scintillator is the standard detection medium for reactor \nuebar{} detectors due to the high abundance of free proton targets, providing excellent pulse shape discrimination, high light yield, and lower cost than plastic scintillator. Liquid scintillators are frequently loaded with gadolinium to decrease the mean neutron capture time and yield a high-energy capture signal. However, gadolinium is ill-suited to compact detectors where neutron-capture $\gamma$-rays will often escape the active volume.
Neutron capture on $^6$Li produces a highly localized ($\ll\!1$~mm) energy deposition from the reaction $n + ^6$Li$ \rightarrow \alpha + t + 4.78$~MeV, making it well suited for use in compact detectors.
While the high $dE/dx$ of the alpha and triton results in a quenched light yield (electron equivalent energy of $0.6$~MeV$_{ee}$), it also allows discrimination from equivalent-energy electromagnetic backgrounds using the PSD capability of certain liquid scintillators. 

\begin{figure}
\centering
\includegraphics[trim={0.cm 0.cm 0cm 0.cm},clip,width=0.34\textwidth]{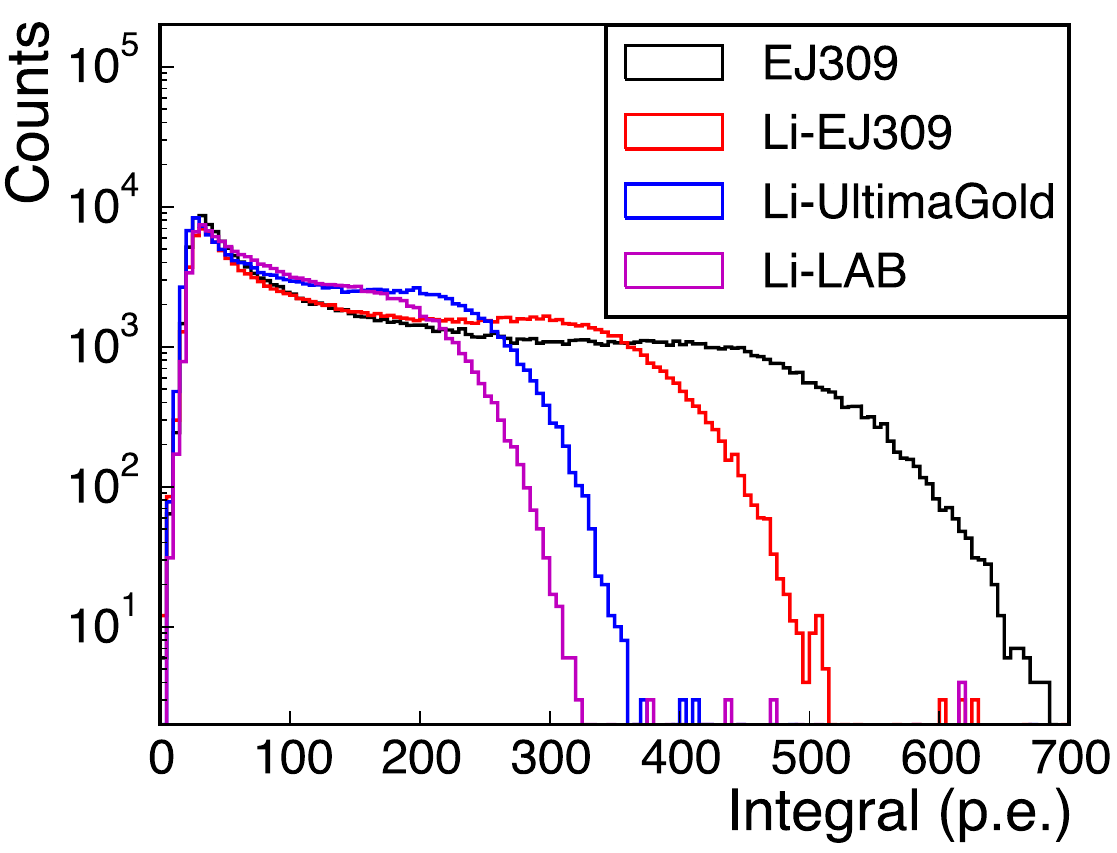} \\
\includegraphics[trim={0.cm 0.cm 0cm 0.cm},clip,width=0.34\textwidth]{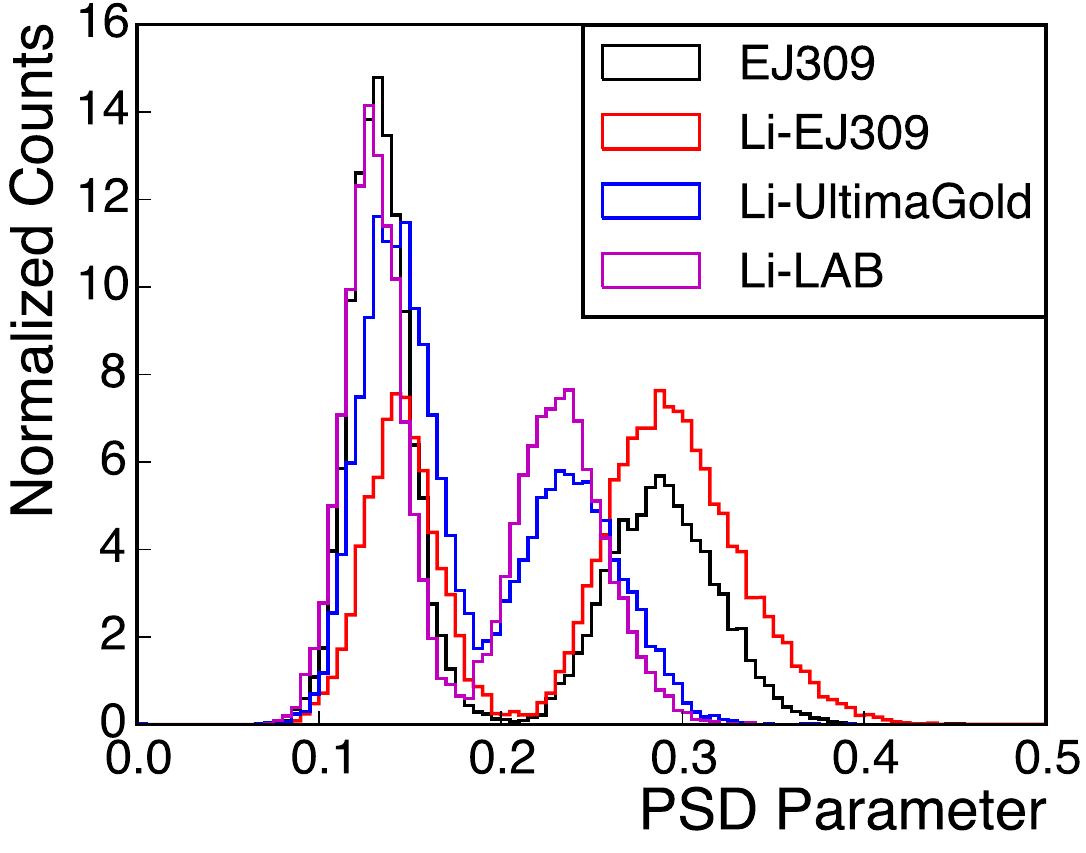} 
   \caption{Comparison between unloaded EJ-309 and three different LiLS formulations. (Top) Response to $^{60}$Co, demonstrating the relative light yield. (Bottom) Comparison of PSD distributions when exposed to $^{252}$Cf. Li-EJ309 has the best performance amongst Li-loaded materials.}
  \label{fig:vialTests}  
\end{figure}

Previously available Li-loaded liquid scintillators were based on toxic and flammable solvents that can no longer be used in reactor facilities.
Therefore, a multiyear research and development effort has explored the feasibility of three new low-toxicity and high-flashpoint scintillator bases, LAB, UltimaGold, and EJ-309~\cite{ugab, ej309}.
Surfactants are used to form a micro-emulsion containing $^6$LiCl, creating a dynamically-stable mixture that retains the PSD capability of the base scintillator.
Extensive studies were performed with each formulation to characterize light yield and PSD performance using $\gamma$-ray and neutron sources.

The EJ-309-based LiLS was found to have the best light yield and PSD performance (Fig.~\ref{fig:vialTests}).  
Li-EJ309 has a proton density of $5.5\times10^{22}$/cm$^3$, light yield above 6500~photons/MeV, and a bulk attenuation length of $\sim$4~m. 
The stability of Li-EJ309 samples has been monitored for approximately one year, with the light yield shown to be stable within the 2\% measurement uncertainty.  

A materials compatibility program studied all components potentially in contact with Li-EJ309 for extended periods of time.
The \adone{} interior will be constructed only with components that have been qualified, i.e. found to be stable and not degrade LS performance.
Acceptable materials include cast acrylic, Teflon, polycarbonate and PLA plastics (clear and colored), Viton, and Acetal. 

\subsubsection{Detector Prototyping}
\label{sec:p20offsite}

\begin{figure}[tbp]

\centering
\includegraphics[trim={0cm 0cm 0cm 0cm},clip,width=0.3\textwidth]{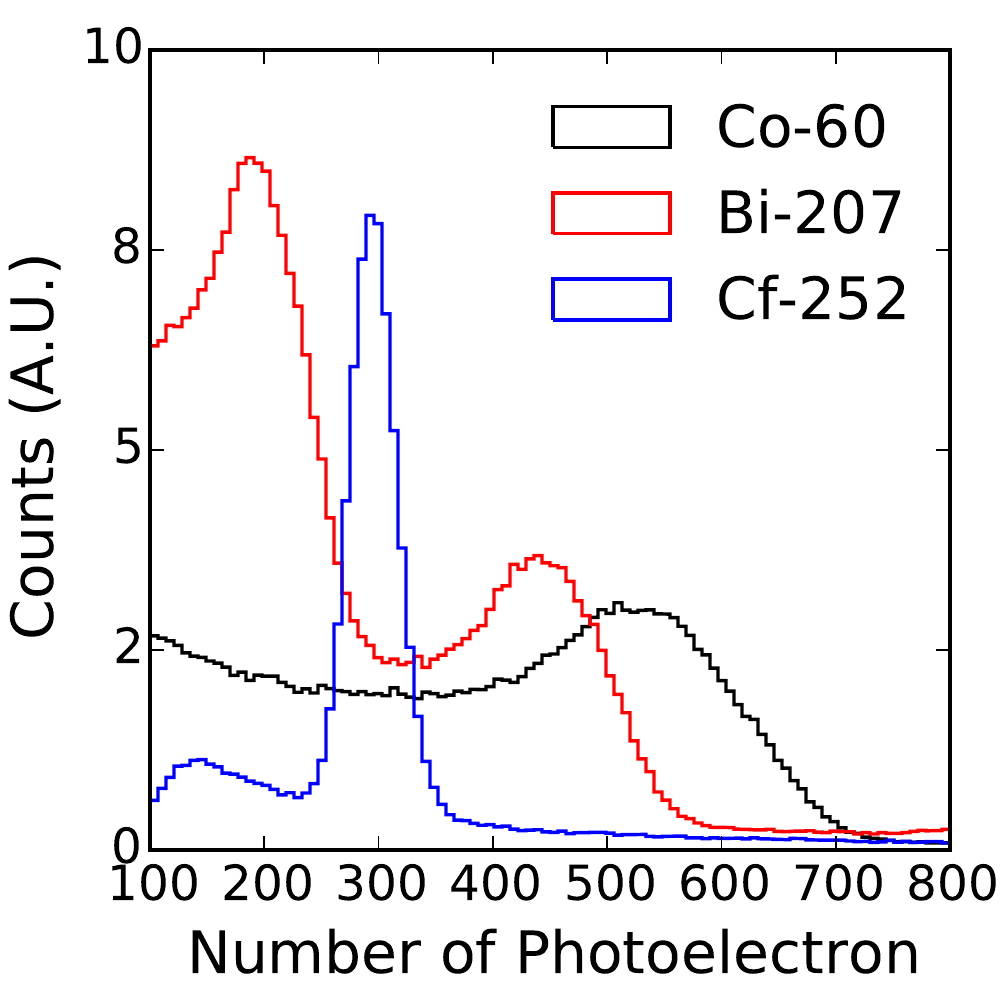}
\includegraphics[trim={0cm 0cm 0cm 0cm},clip,width=0.3\textwidth]{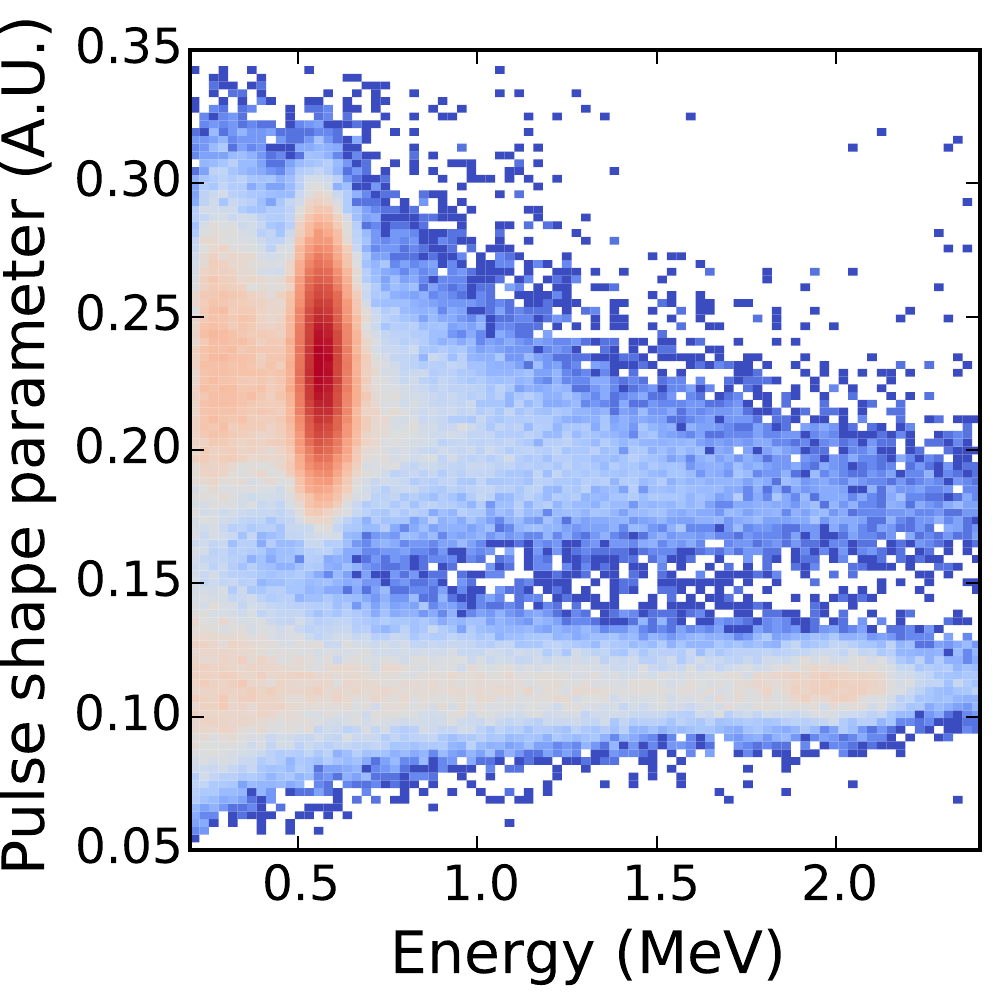}
\caption{(Top) Measured PE spectra including the Compton edge of $^{60}$Co and $^{217}$Bi $\gamma$-rays and the quenched ($n$, Li) capture peak from $^{252}$Cf neutron source. (Bottom) PSD performance of Li-EJ309. The upper band is neutron-like events with ($n$,Li) captures at $\approx0.6$~MeV dominating the statistics. The lower band shows $\gamma$-like events.  }
  \label{fig:p20_data}  
\end{figure}

\pspt{} has prototyped many key elements of the proposed \adone{} design, including constructing test detectors to validate the light collection efficiency and PSD performance and the low-mass reflector system that optically segments the active liquid scintillator target.

\begin{figure*}[t]
\centering
\includegraphics[trim=0cm 0cm 0cm 0cm, clip=true, width=0.80\textwidth]{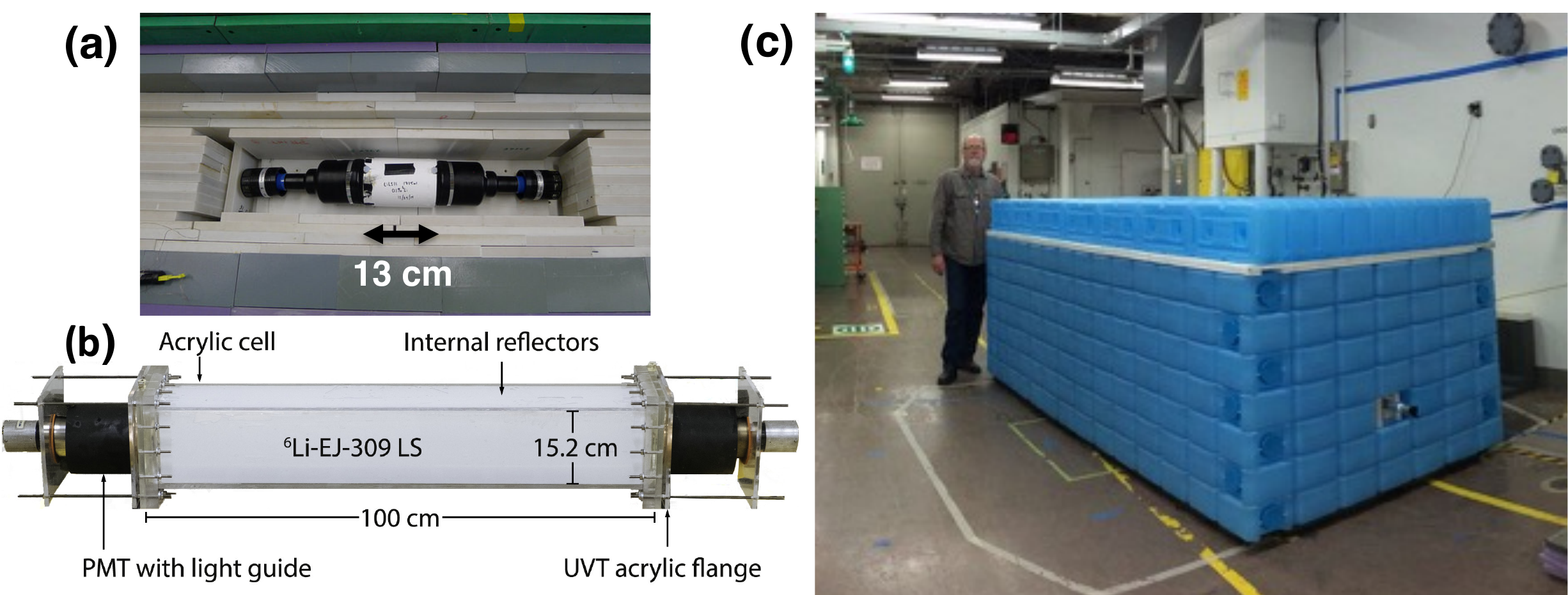}
 \caption{Prototype detectors and shielding installed at the HFIR experimental location. (a) The PROSPECT-2 prototype within a partially assembled polyethylene and lead shield. (b) The PROSPECT-20 prototype with internal reflectors added after operation at HFIR.  (c) The PROSPECT-20 shielding enclosure at HFIR.}
\label{fig:onsitePrototypes}
\end{figure*}

An acrylic test detector (15 cm $\times$ 15 cm $\times$ 1 m), referred to as "PROSPECT-20", was produced to validate the performance of the \adone{} optical design.
The effects of different PMT models, single versus double-ended readout, reflector types and coupling methods have been explored and reported in~\cite{p20paper}. 
As shown in Fig.~\ref{fig:onsitePrototypes}, the detector utilized internal reflectors similar to the low-mass panels discussed below and the R6594 PMTs chosen for \adone{}~\cite{R6594}.  
Filled with EJ-309~\cite{ej309}, a light collection of 841$\pm$17 photoelectrons (PE)/MeV was observed with excellent PSD performance: a rejection factor of $10^4$ for $\gamma$-rays was achieved while preserving 99.9\% of the ($n$,Li) capture signal between 0.5--0.7~MeV.  In addition, both PSD and light collection were found to be totally uniform along the length of the cell within systematic and statistical uncertainties.  
Position reconstruction along the long axis of PROSPECT-20 utilizing light arrivial time differences between PMTs was also demonstrated with a 5~cm resolution.

When filled with LiLS an average light collection of 522$\pm$16 PE/MeV was measured with three $\gamma$-ray sources in the range 0.38--2.0 MeV (Fig.~\ref{fig:p20_data}).
The demonstrated PE/MeV exceeds the goal of 500 PE/MeV needed to achieve the target 4.5\%/$\sqrt{E({\rm MeV})}$ energy resolution. 
Excellent uniformity and PSD performance was again demonstrated at the ($n$,Li) capture peak and above (Fig.~\ref{fig:p20_data})  
enabling preservation of 99.9\% of the ($n$,Li) signal while rejecting the same fraction of $\gamma$-ray  events.
The mean neutron capture time is observed to be 40~$\mu$s.

For the optical segmentation system, low-mass reflector panels have been developed by adhering 3M Enhanced Specular Reflector (ESR)~\cite{esr} to both sides of a rigid $0.6$~mm thick carbon fiber sheet, and then enclosing this assembly in a Teflon sleeve via heat bonding. The result is a large area, low-mass, highly reflective assembly that is %fully 
hermetic and liquid scintillator-compatible.
A structural support system, described in Sec.~\ref{sec:detectorDesign}, was prototyped using polycarbonate and 3D-printed white PLA plastic, which allowed for rapid fabrication and testing cycles. 
A nine-segment mechanical mock-up of the structure has been used to develop assembly procedures and demonstrate the mechanical robustness of the segmentation system. 

\subsubsection{Onsite Prototype Detector Operation and Simulation Validation}
\label{sec:p20onsite}

PROSPECT has deployed multiple liquid scintillator prototype detectors and shielding packages at HFIR since mid-2014 to characterize backgrounds \textit{in-situ} and develop a working knowledge of facility regulations, operating procedures and work control processes. 
Detector size was increased by a hundredfold, from an initial 100~ml EJ-309 cell to a 23~l cell containing LiLS (PROSPECT-20). 
The shielding packages have likewise grown from a small lead brick cave to a multilayered shield of water bricks~\cite{Waterbricks}, High Density Polyethylene (HDPE), 5\% borated HDPE and lead with a total volume nearly 1/4 that of the proposed \adone{} design (Fig.~\ref{fig:onsitePrototypes}).

\begin{figure}[tbp]
\centering
    \includegraphics[trim={0cm 0cm 0cm 0cm},clip,width=0.35\textwidth]{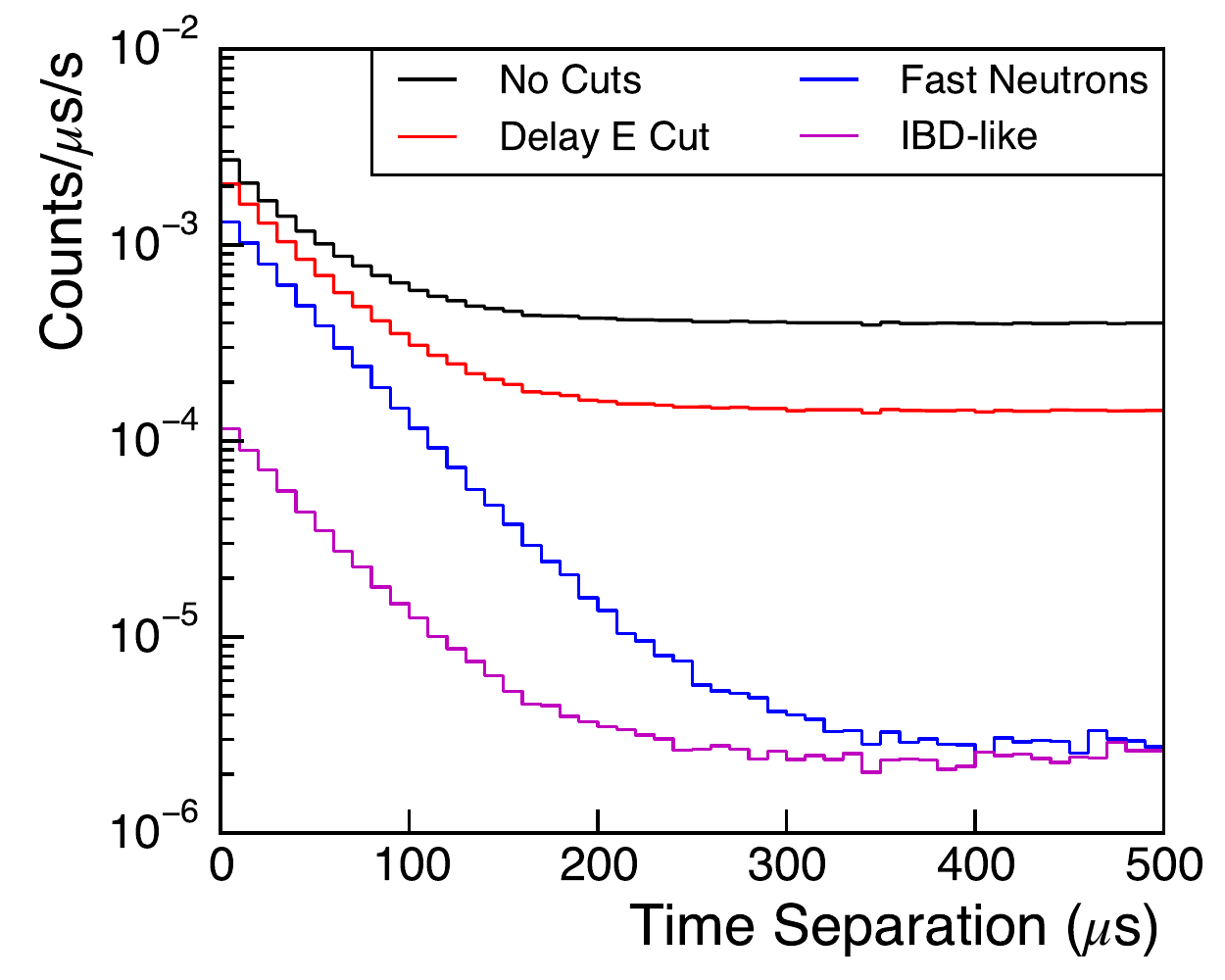}
    \includegraphics[trim={0cm 0cm 0cm 0cm},clip,width=0.35\textwidth]{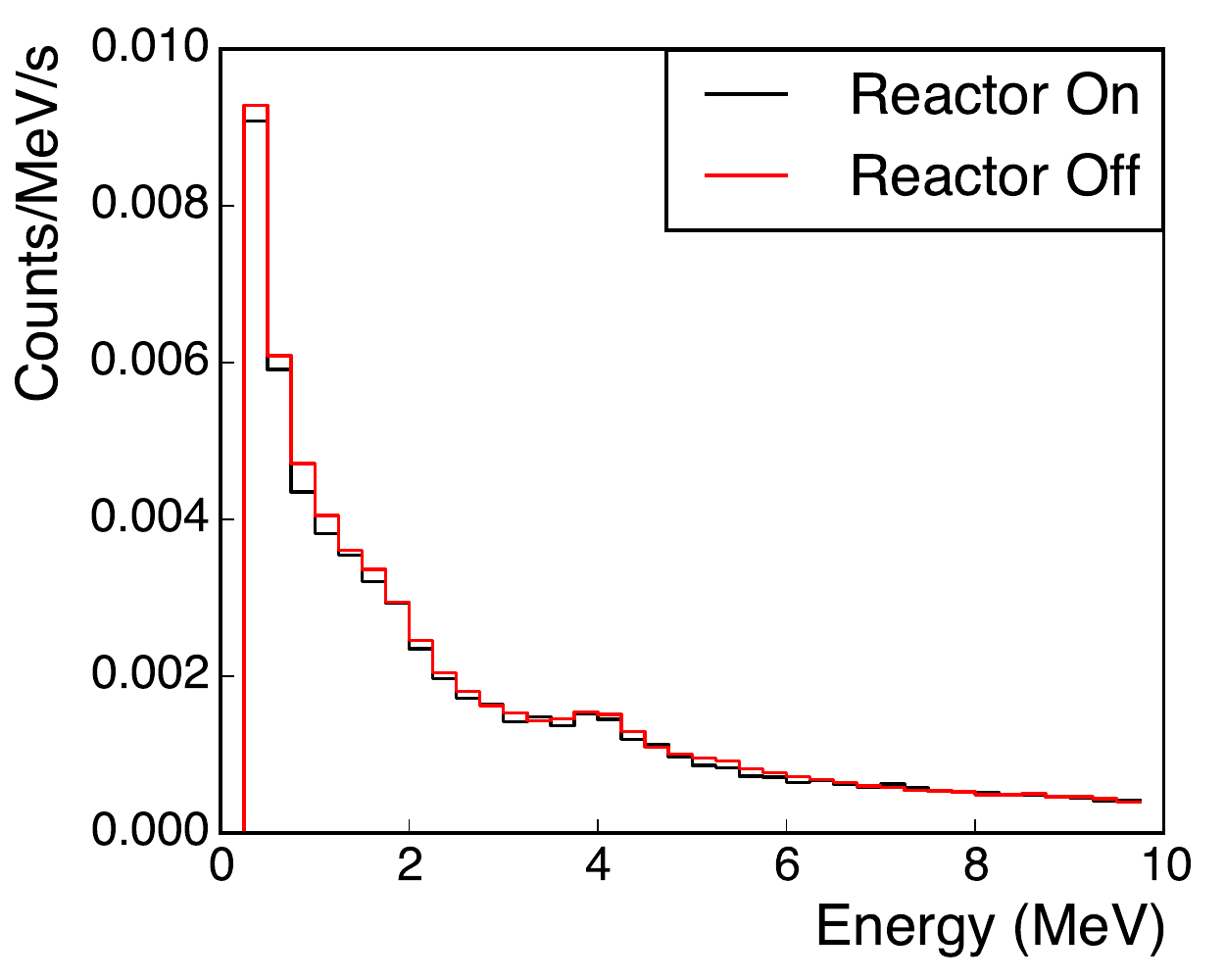}

    \caption{(Top) Time separation between prompt and delayed signals (black) for reactor-off operation of PROSPECT-20 with different analysis cuts applied: delay energy (red), fast neutrons (blue), and IBD-like (purple). %: e$^+$-like followed by $^6$Li neutron capture. 
   (Bottom) Comparison of PROSPECT-20 IBD-like event prompt energy spectra with HFIR on (black) and off (red).   }
  \label{fig:p20_HFIR_data_sim_spec_comp}
\end{figure}

Data have been collected over nine months through multiple reactor cycles. 
Analysis cuts were developed to isolate IBD-like events and elucidate the event types that produce background at this location. 
The time separation (Fig.~\ref{fig:p20_HFIR_data_sim_spec_comp} top) spectrum between prompt and delayed signals is dominated by a random background that is constant in time, but also exhibits an exponentially decaying time-correlated component consistent with that observed for correlated particle processes that terminate in a neutron capture. IBD is one such process (prompt $e^+$ followed by neutron capture), as are correlated backgrounds due to fast neutron recoil followed by capture or capture of multiple spallation neutrons.

Application of a simple energy cut around the ($n$,Li) capture peak for the delayed signal in an event pair reduces the random component by a factor of 2.8, demonstrating, in part, the utility of LiLS for a compact detector. 
Applying selections based upon PSD provides further information: requiring that the prompt signal fall in the neutron recoil band (blue curve in Fig~\ref{fig:p20_HFIR_data_sim_spec_comp}) indicates that the majority of time-correlated background events in PROSPECT-20 are due to fast neutron recoil followed by capture. 
Finally, applying selections consistent with IBD events (prompt PSD in $\gamma$-like band, delayed signal in ($n$,Li) energy and PSD region) reduces the initial coincidence rate by a factor of 55 and reveals the IBD-like background to be dominated by time-correlated pairs (magenta curve).  
Accidental coincidences due to reactor-produced $\gamma$-rays following this selection are minimal due to the selectivity of the $^6$Li neutron capture signature and targeted shielding applied to background ``hot-spots'' at HFIR. 
Comparison of IBD-like event energy spectra with the reactor  on and off (Fig.~\ref{fig:p20_HFIR_data_sim_spec_comp} bottom) indicates that IBD-like backgrounds are cosmogenic and that reactor generated backgrounds are negligible.

% !TEX root = main.tex

\subsection{Timeline}
\label{sec:timeline}

PROSPECT Phase~I will be ready to proceed in early 2016. 
Design, construction, assembly, and installation of \adone{} will take approximately one year.  
Installation of components in the HFIR experimental room is restricted to the reactor-on periods due to limited availability of HFIR craft labor during reactor shutdowns and maintenance periods. 
Detector assembly and installation will be completed by the end of calendar year 2016. 
After a brief period of commissioning, data taking will start in early 2017.
\adone{} will run continuously for three years over all reactor-on/off cycles.  
Periodic calibrations are planned and will minimally interrupt continuous data taking. 
At least three detector movements to different baselines are planned. Since \adone{} and the associated shielding are mounted on air bearings, no disassembly is required.  
Detector repositioning and re-start  will take $\leq$2 days and will have minimal impact on data collection. 
With timely funding, construction of Phase~II could start soon, with data taking about 2--3 years later.

% !TEX root = main.tex

\section{Predicted Detector Response: Signal and Background}
\label{sec:bkg}

A comprehensive and flexible Monte Carlo simulation of the \pspt{} detector design has been developed using the \textsc{Geant}4 package \cite{Agostinelli:2002hh}.
Particle interactions are based on the ``\texttt{QGSP\_BERT\_HP}'' physics list in \textsc{Geant}4.10.01p1, which focuses on
	``high precision'' models for lower-energy neutron interactions.
Optical photon generation and tracking is optionally available for light transport modeling.
The simulation includes geometries for prototype test detectors and the two ADs.
A variety of event generators are available, including inverse beta decay,
	\texttt{CRYv1.7} \cite{Hagmann2007} for cosmic ray shower generation,
	a parametrized model for the surface cosmic neutron spectrum \cite{Sato2006544},
	and calibration sources.

\subsection{Response to the Reactor \nuebar{} Signal}
\label{sec:sigResponse}

Simulation studies have been used to study the response of the \pspt{} \adone{} design to the inverse beta decay signal and many classes of background events. This includes particle transport studies using \textsc{Geant}4 and examination of  segment optical response using both \textsc{Geant}4 and the SLitrani package~\cite{SLitrani}. Exploratory studies of segment response confirmed that the choice of an efficient specular reflector, with a component of total internal reflection (TIR) from the Teflon layer encapsulating the segment wall, provided good collection efficiency and uniformity along the entire segment volume. Good agreement was found with the data from the \ptwenty{} detector using various reflector and light guide configurations.

\begin{figure}[tbp]
\centering
\includegraphics[clip=true, trim=0mm 0mm 0mm 10mm,width=0.3\textwidth]{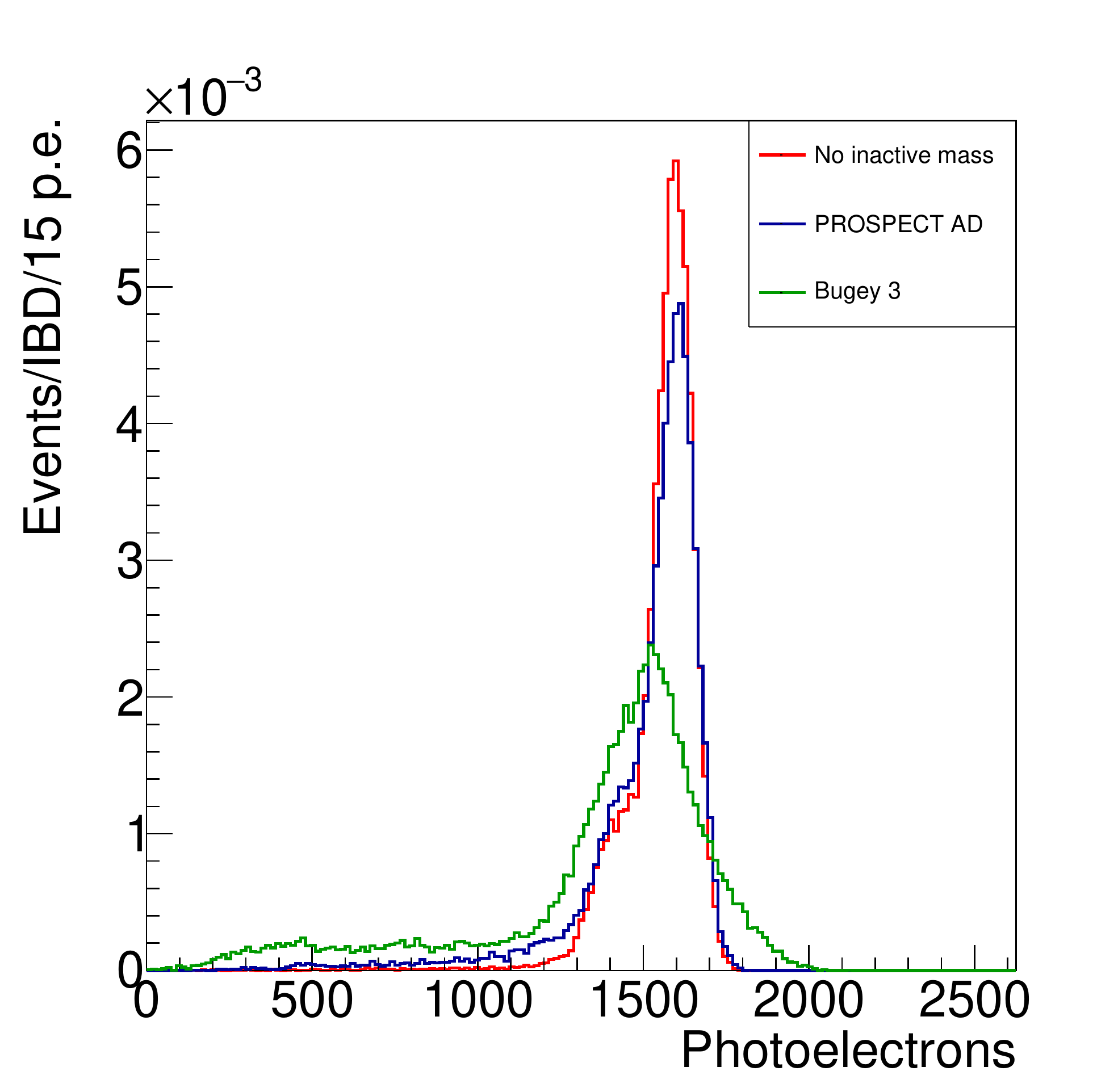}\hfil
 \caption{ Simulated response to 4~MeV e$^+$  for \adone{} configurations with no inactive mass (red), the \pspt{} low mass optical separators (blue), and an inactive mass fraction equivalent to Bugey~3 (green).}
\label{fig:wallStudy}
\end{figure}

The response to the $e^+$ produced by IBD  is of particular interest and has been examined for a wide range of segment size, segment wall compositions, and optical configurations. The $e^+$ response achieved with the \adone{} design described above is illustrated in  Fig.~\ref{fig:wallStudy}, which includes both particle and optical scintillation photon transport. Here, the response to mono-energetic $4$~MeV positrons distributed uniformly throughout the fiducial detector volume is examined for several segment wall configurations: walls with infintesimal thickness (no inactive mass in active volume), the \adone{} design, and a thickness equivalent to that used in the  segmented Bugey~3 \nuebar{} detector~\cite{Abbes:1995nc}. The \adone{} configuration of low-mass separators has comparable performance to a detector without inactive material and is significantly better than the  segmented Bugey~3 detector.

\subsection{Backgrounds from Cosmogenic Activity}
\label{sec:CosmicBG}

\begin{figure}[tbp]
\centering
\includegraphics[width=0.3\textwidth]{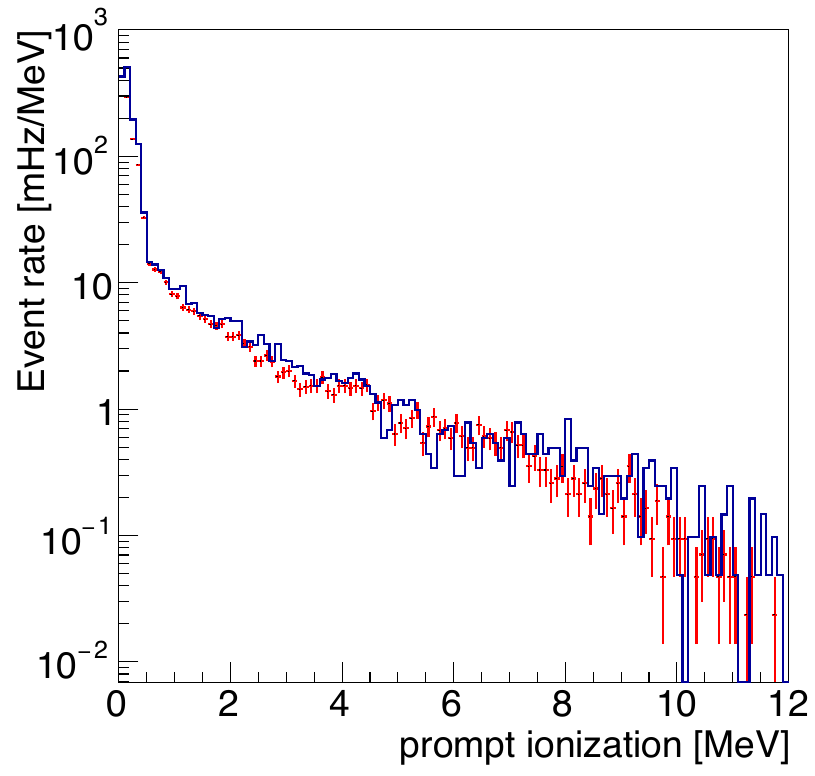}
\includegraphics[width=0.3\textwidth]{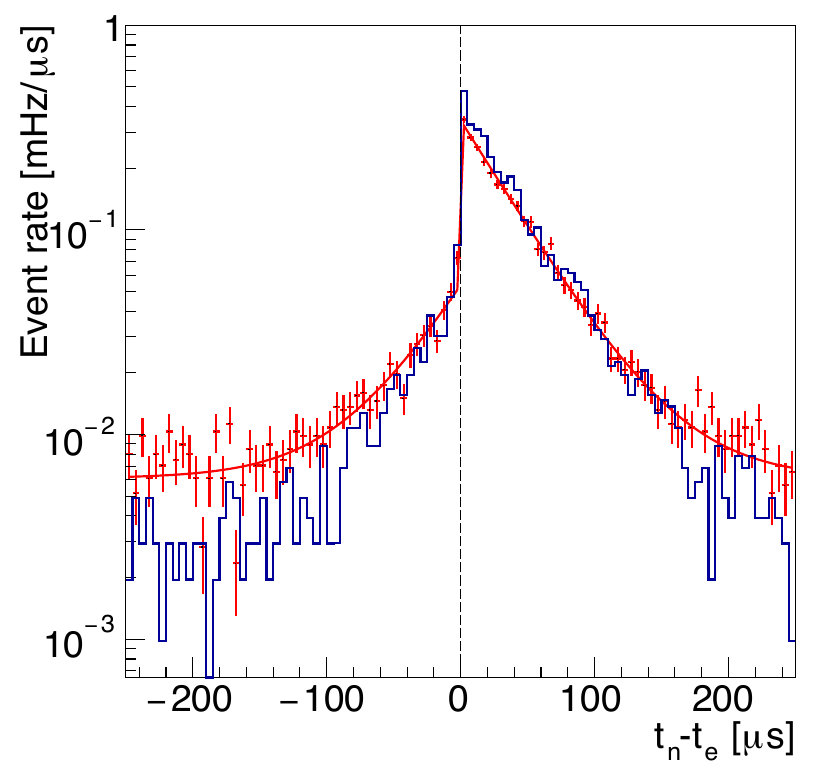}
  \vspace{-7pt}
   \caption{(Top) PROSPECT-20 IBD-like energy distributions from reactor-off data (red) compared to simulation for cosmic backgrounds (blue). (Bottom) PROSPECT-20 IBD-like timing distributions from reactor-off data (red) compared to simulation for cosmic backgrounds (blue). }
  \label{fig:p20_HFIR_data_sim_time_comp}  

\end{figure}

Data collected using the PROSPECT-20 detector at HFIR have been used to validate the \pspt{} \adone{} simulation.
For example, Fig.~\ref{fig:p20_HFIR_data_sim_time_comp} displays an absolute comparison between data and simulation of cosmic ray shower backgrounds.
Both the energy and time distributions of IBD-like events are in good agreement, with the results being consistent with fully explaining the observed IBD-like rate in PROSPECT-20.
Although the IBD-like background rate is higher than the expected \nuebar{} interaction rate,
	improved shielding and cuts possible in the full \adone{} will suppress backgrounds substantially,
	achieving a signal-to-background ratio of $\geq\!1$ according to simulation.

\begin{figure}[tbp]
\centering
\includegraphics[clip=true, trim=0mm 0mm 0mm 0mm,width=0.30\textwidth]{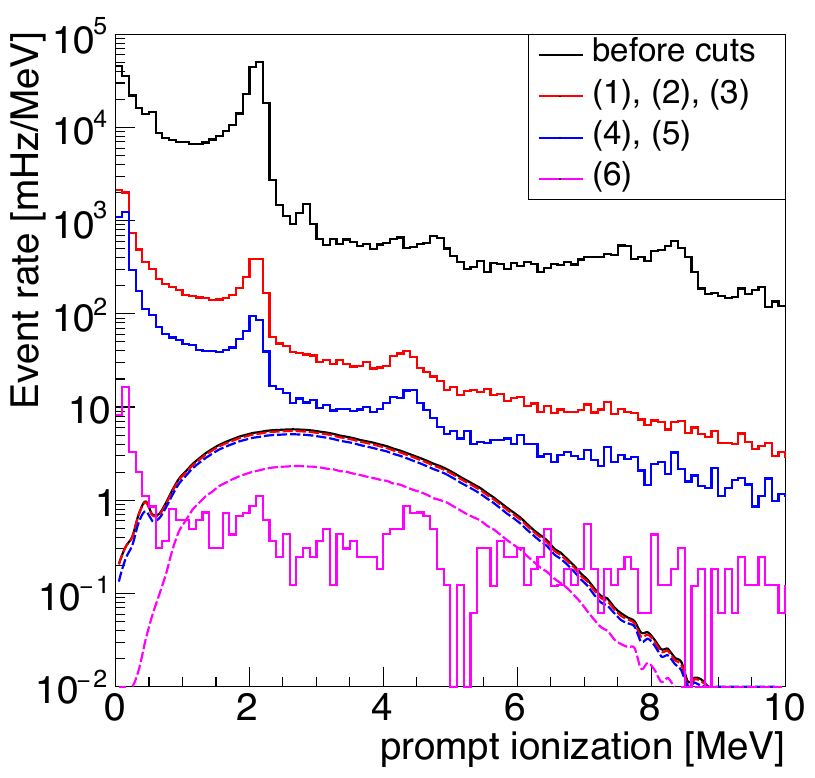}
\includegraphics[width=0.30\textwidth]{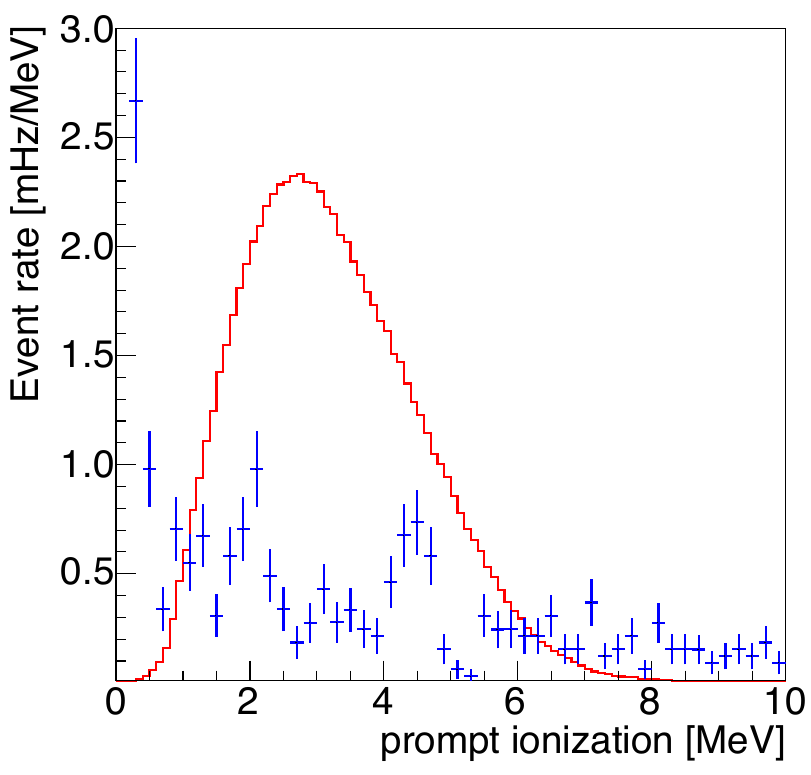}\hfil
 \caption{(Top) Simulated \adone{} IBD signal and background spectra. Signal (dashed) and background (solid) prompt spectra are shown through selection cuts described in the text. Background is primarily produced by cosmogenic fast neutrons. (Bottom) Simulated \adone{} IBD signal (red) and background (blue) spectra after all analysis cuts. Signal-to-background of better than 1:1 is predicted. The rate and shape of the residual IBD-like background can be measured with high precision during reactor off periods.}

\label{fig:cosmic_IBD_spectrum_cuts}
\end{figure}

The validated simulation package indicates that measured IBD-like background in PROSPECT-20 is  primarily due to high-energy (tens to hundreds of MeV) cosmic neutrons, with small additions from muon interactions and accidental $\gamma$-ray coincidences.
These mechanisms are also projected to be the primary source of IBD-like background events in the \pspt{} \adone{}.
By design, the multi-segment \adone{} provides information useful for identifying and vetoing most cosmic background events.
However, high-energy cosmic neutrons, which can penetrate undetected deeply into the active volume before inelastic scattering interactions, can produce time-correlated prompt ionization, highly quenched nuclear recoils, and delayed secondary neutron capture signals, and are projected to be the main background source.
The rates of cosmogenically-produced $^9$Li and $^8$He, which also mimic IBD signals, are estimated to be roughly two orders of magnitude below the IBD rate, and can be measured with reactor-off data. % is this detectable under other backgrounds? --MPM

After identification of candidate prompt and delayed signals via deposited energy and PSD selections, additional cuts on event topology (including both time and position information) provide two to three orders of magnitude in background suppression.
Fig.~\ref{fig:cosmic_IBD_spectrum_cuts} demonstrates the effectiveness of topology cuts at rejecting cosmic ray background relative to the IBD signal. 
%\ref{sec:oscSens}.

\begin{table}[tbp]
\centering
\small
\begin{tabular}{| l | c | c |} \hline
   \textbf{Cuts} & \textbf{IBD signal} & \textbf{Cosmic BG} \\ \hline
   \textbf{PSD} 			& 1630 	& 2.1e6 \\ \hline
   \textbf{Time (1, 2, 3)}		& 1570	& 3.4e4 \\ \hline
   \textbf{Spatial (4, 5)}		& 1440	& 9900 \\ \hline
   \textbf{Fiducial (6)}		& 660	& 250 \\ \hline
\end{tabular}
  \caption{Simulated signal and cosmic background rates in events per day in the energy range $0.8 \leq E \leq 7.2$\,MeV, after applying background rejection cuts.}
  \label{tab:bg_cuts}
\end{table}

The event selections  are as follows. ``Time topology'' cuts include:
(1) delayed capture must occur within 100~$\mu$s of the prompt ionization; 
(2) multiple hits in the prompt cluster must occur within 5~ns to reject slower-moving neutron recoil events;
(3) events must be isolated from other neutron recoils or captures in a $\pm 250\,\mu$s window, to reject multi-neutron spallation showers.
``Spatial topology'' cuts include: 
(4) the prompt and delayed signals must be proximate; 
(5) multiple segment hits in the prompt signal must be distributed over a compact volume, rejecting extended minimum ionizing tracks and many high-energy gammas;
(6) events occurring outside the inner fiducial volume ($\geq$ one segment width from any active volume surface) are vetoed.

Although fiducialization decreases the effective active volume for true IBDs by $\sim$50\%, it provides a more than tenfold boost in background rejection.
Predicted rates with cuts are given in \autoref{tab:bg_cuts}.
Within the fiducialized volume, IBD detection efficiency is 42\%.
Efficiencies for the fiducial segments are largely consistent: a percent-level 1$\sigma$ deviation in efficiencies between segments can be corrected for in an oscillation analysis utilizing gamma and neutron calibration results.
Monte Carlo investigations have also characterized the small expected geometry-induced spectral differences between segments within the fiducialized volume.  

While event selection is not yet optimized, the signal-to-background ratio indicated is more than sufficient to meet the \pspt{} physics goals (Sec.~\ref{sec:oscSens}). 
Although the background prediction is inherently uncertain (estimated to be accurate within a factor of 2), we are confident likely gains from optimization will ensure that \pspt{} can successfully control backgrounds.  
The simulated background spectrum shown in Fig.~\ref{fig:cosmic_IBD_spectrum_cuts}(b) is used in the sensitivity calculations described in Sec.~\ref{sec:physics-program}.  
Reactor-off periods will be used to measure the background spectrum directly. 

\subsection{Backgrounds from Internal Radioactivity}
\label{sec:internalBG}

Internal contamination of trace radioactivity can introduce backgrounds that cannot be effectively removed through fiducialization.
These predominantly consist of $^{40}$K and the uranium and thorium decay chains. 
Work has been done to identify radio-clean materials well-suited for installation in the detector package.
Acrylic and other plastic polymers have largely been demonstrated to be low-background by many experiments, including Daya Bay and other $\theta_{13}$ experiments.
The internal reflectors are designed to be low-mass and will mainly consist of carbon fiber, another established low-background material.
Radioactivity from the PMTs will be mitigated by a combination of passively moderating light guides and active fiducialization along a segment's length.

\begin{figure}[tbp]
\begin{center}
\includegraphics[width=.4\textwidth]{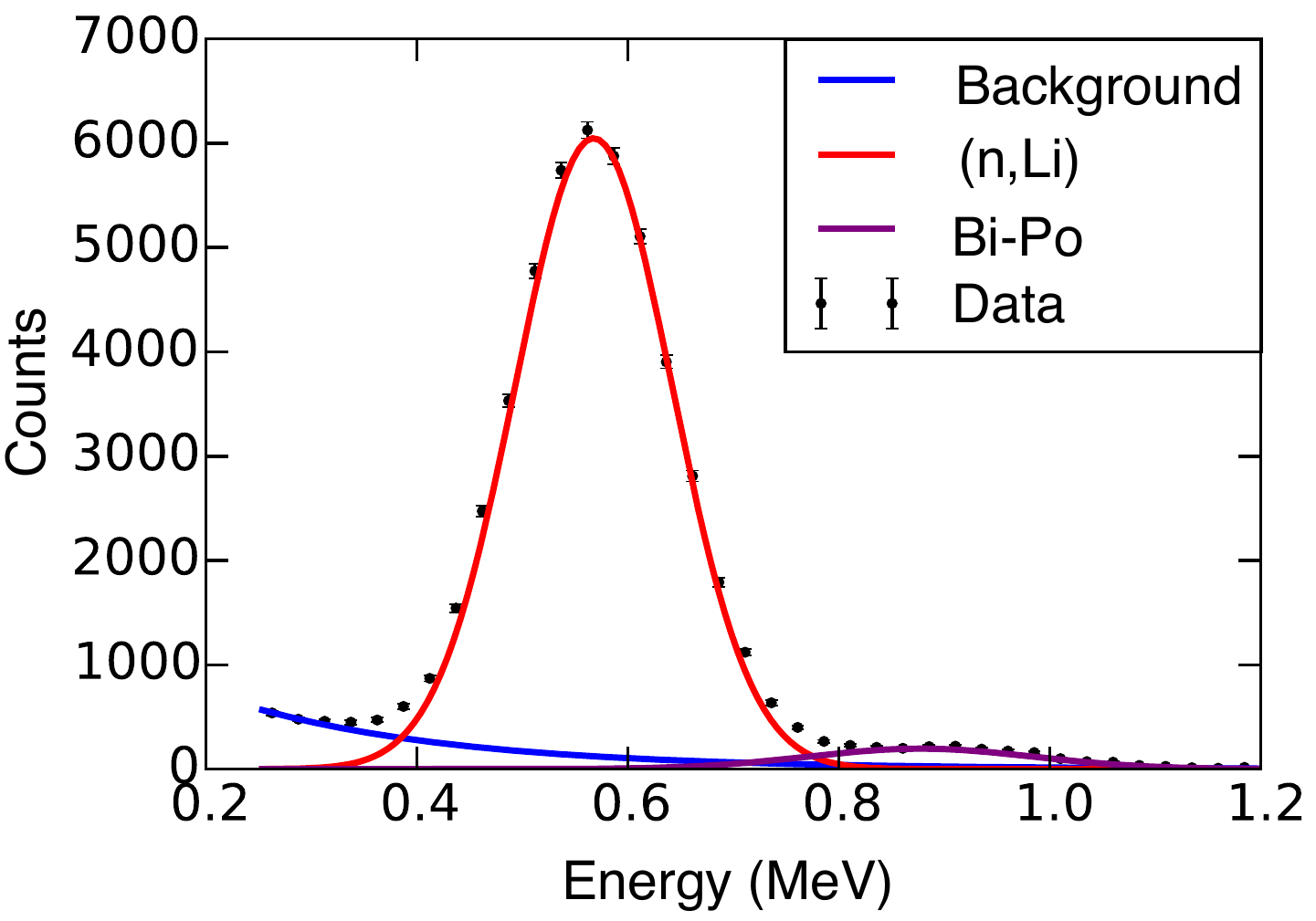}
\caption{Fitted distribution of delayed energy depositions in the PROSPECT-20 detector collected at HFIR. The main peak is from neutron captures on $^6$Li, while the smaller peak at 0.89~MeV is from Bi-Po alpha decays. With demonstrated improvements in energy resolution, these peaks will be completely separated in \pspt{}}
\label{fig:P20bipo}
\end{center}
\end{figure}

Few decays produce correlated signals that can mimic the IBD event signature: electron-recoil like prompt event with a delayed neutron capture. 
However, correlated $^{214}$Bi $\rightarrow$ $^{214}$Po $\rightarrow$ $^{210}$Pb decays in the $^{238}$U decay chain are one of them. 
Bi-Po decays consist of a beta-decay with endpoint of approximately 3~MeV followed by a 7.8~MeV alpha decay with a half-life of 164~$\mu$s.  
Data collected with PROSPECT-20 at HFIR, shown in Fig.~\ref{fig:P20bipo}, demonstrate how Bi-Po decays appear in the \pspt{} data stream.

A rate of Bi-Po events of $3.0\pm0.15$/day/liter was observed with the delayed alpha's quenched light output of 0.89~MeV$_{ee}$. 
With the demonstrated energy resolution of the internal-reflector phase of PROSPECT-20, the alpha peak will be well-separated from the (n,Li) capture peak, eliminating greater than 99.5\% of the Bi-Po events while preserving greater than 99\% of neutron captures.  
Applying this rejection power, the Bi-Po signal is reduced to approximately 30/day, an order of magnitude below the IBD rate.

% !TEX root = main.tex

\section{Conclusions}
\label{sec:conclusions}
%\graphicspath{ {./figs/}

Since their first observation some 60 years ago, reactor antineutrinos have been an excellent tool for the study of neutrinos and neutrino oscillation. As an experiment at very short baselines from a research reactor with a highly-enriched uranium core, PROSPECT will make a precise measurement of the reactor antineutrino spectrum from $^{235}$U with an energy resolution of better than 4.5\%  at the High Flux Isotope Reactor at ORNL and search for neutrino oscillations as a sign of eV-scale sterile neutrinos. Utilizing a single, segmented 3-ton liquid scintillator detector located at 7--12~m from the reactor core, PROSPECT Phase~I will probe the favored region of parameter space at $>$3$\sigma$ within 3 years of data taking. Phase~II will add a second detector with $\sim$10-tons active target mass and cover the majority of the allowed parameter space at $>$5$\sigma$. PROSPECT has performed extensive R\&D on detector components and built and operated a series of test detectors with increasing size. These test detectors have allowed the characterization and validation of the detector design as well as background studies in the reactor environment at HFIR. PROSPECT is technically ready to proceed with the construction of Phase~I of the experiment and a first physics result can be obtained within 1 year of data taking. 

% !TEX root = main.tex

\section{Acknowledgements}
\label{sec:acknowledgements}

This material is based upon work supported by the U.S. Department of Energy Office of Science. 
Additional support for this work is provided by Yale University, the Illinois Institute of Technology, the National Institute of Standards and Technology, and the Lawrence Livermore National Laboratory LDRD program. 
We gratefully acknowledge the support and hospitality of the High Flux Isotope Reactor and the Physics Division at Oak Ridge National Laboratory, managed by UT-Battelle for the U.S. Department of Energy.

\bibliography{bib-PROSPECT-physics}

\end{document}